\documentclass[11pt,twocolumn]{IEEEtran}
\usepackage[dvips]{graphicx}
\usepackage{rotating}
\newif\ifremark
\long\def\remark#1{
\ifremark%
        \begingroup%
        \dimen0=\textwidth
        \advance\dimen0 by -1in%
        \setbox0=\hbox{\parbox[b]{\dimen0}{\protect\em #1}}
        \dimen1=\ht0\advance\dimen1 by 2pt%
        \dimen2=\dp0\advance\dimen2 by 2pt%
        \vskip 0.25pt%
        \hbox to \textwidth{%
                \vrule height\dimen1 width 3pt depth\dimen2%
                \hss\copy0\hss%
                \vrule height\dimen1 width 3pt depth\dimen2%
        }
        \endgroup%
\fi}
\remarktrue

\begin{document}

\title{Small-World File-Sharing Communities}

\author{Adriana Iamnitchi, Matei Ripeanu, Ian Foster\\
\textit{Department of Computer Science}\\ 
\textit{The University of Chicago}\\
\textit{Chicago, IL 60637}\\
\textit{\{anda, matei, foster\}@cs.uchicago.edu}
}

\date{}
\maketitle
\thispagestyle{empty}
\begin{abstract}

Web caches, content distribution networks, peer-to-peer file sharing
networks, distributed file systems, and data grids all have in common
that they involve a community of users who generate requests for
shared data. In each case, overall system performance can be improved
significantly if we can first identify and then exploit interesting
structure within a community's access patterns. To this end, we
propose a novel perspective on file sharing based on the study of the
relationships that form among users based on the files in which they
are interested.
We propose a new structure that captures common user interests in
data---the \textit{data-sharing graph}--- and justify its utility with
studies on three data-distribution 
systems: a high-energy physics collaboration, the Web, and the Kazaa
peer-to-peer network. We find small-world patterns in the data-sharing
graphs of all three communities. We analyze these graphs and propose
some probable causes for these emergent small-world patterns. The
significance of small-world patterns is twofold: it provides a
rigorous support to intuition and, perhaps most importantly, it
suggests ways to design mechanisms that exploit these naturally
emerging patterns.

\end{abstract}

\section{Introduction}
Large-scale, Internet-connected distributed systems are notoriously
difficult to manage. In a resource-sharing environment such as a
peer-to-peer system that connects hundreds of thousands of computers
in an ad-hoc network, intermittent resource
participation, large and variable scale, and high failure rates are
challenges that often impose performance tradeoffs. Thus,
existing P2P file-location mechanisms favor specific requirements: in
Gnutella, the emphasis is on accommodating highly volatile peers and 
on fast file retrieval, 
with no guarantees that files will always be located. In Freenet \cite{clarke00freenet}, the
emphasis is on ensuring anonymity. In contrast, distributed hash
tables such as CAN \cite{ratnasamy01can}, Chord \cite{stoica01chord}, Pastry \cite{rowstron01pastry},
and Tapestry \cite{zhao01tapestry} guarantee that files 
will always be located, but do not support wildcard searches. 

One way to optimize these tradeoffs is to understand user
behavior. In this paper we analyze user behavior in three file-sharing
communities in an attempt to get inspiration for designing efficient
mechanisms for large-scale, dynamic, self-organizing resource-sharing
communities.  

We look at these communities in a novel way: we study the
relationships that form among users based on the 
data in which they are interested. We capture and quantify these
relationships by modeling the community as 
a \textit{data-sharing graph}. To this end, we propose a new structure
that captures common user interests in 
data (Section \ref{sec:DSG}) and justify its utility with studies on
three data-distribution systems 
(Section \ref{sec:3communities}): a high-energy physics collaboration,
the Web, and the Kazaa peer-to-peer 
network. We find small-world patterns in the data-sharing graphs of
all three communities 
(Section\ref{sec:SWDSG}). We discuss the causes of these emergent
small-world patterns in Section 
\ref{sec:HNZL}. The significance of these newly uncovered patterns is
twofold (Section \ref{sec:DSG-uses}): 
First, it explains previous results \cite{sripanidkulchai03efficient}
and confirms (with formal support) 
the intuition behind them. Second, it 
suggests ways to design mechanisms that exploit these naturally
emerging patterns.

\section{Intuition}

It is not news that understanding the system properties can help guide
efficient solution design. A well known example is the 
relationship between file popularity in the Web and
cache size. The popularity of web pages has been shown to follow a
Zipf distribution
\cite{barford98changes, 
breslau99web}: few pages are highly
popular and many pages are requested few times. As a result, the
efficiency of 
increasing cache size is not linear: caching is useful for the
popular items, but there is little gain from increasing the cache to
provision for unpopular items.  

As a second example, many real networks are power
law. That is, their node degrees are distributed according 
to a power law, such that a small number of nodes have large degrees,
while most nodes have small degrees. Adamic et
al. \cite{adamic01search} propose a  mechanism for probabilistic
search in power-law networks that exploits exactly this  
characteristic: the search is guided first to nodes with high degree and
their many neighbors. This way, a large percentage of the network is
covered fast. 


This type of observations inspired us to look for patterns in user
resources requests. But what patterns?

\subsection{Patterns, Patterns Everywhere}

It is believed that the study of networks started with Euler's
solution of the K\"{o}nigsberg bridge problem in 1735. The
field has since extended from theoretical results to the analysis of
patterns in real networks. Social sciences have apparently the longest
history in the study of real networks \cite{newman03structure}, with
significant quantitative results dating from the 1920s
\cite{freeman96antecedents}. 

The development of the Internet added significant momentum to
the study of networks: by both facilitating access to
collections of data and by introducing new networks to study, such as
the Web graph, whose nodes are  web pages and edges are hyperlinks 
\cite{broder00web}, the Internet at the router and the AS level
\cite{faloutsos99powerlaw} and the email graph \cite{newman02email}. 

The study of large real networks led to fascinating results:
recurring patterns emerge in real networks (see
\cite{albert02, barabasi02linked, dorogovtsev02evolution,
newman03structure} for good surveys). For example, a frequent pattern
is the power-law  distribution of node degree, that is, a small 
number of nodes act as hubs (having a large degree), while most nodes
have a small degree. Examples of power-law networks are numerous and
from many domains: the phone-call network (long distance phone calls
made during a single day) \cite{abello99maximum, aiello00random}, the
citation network \cite{redner98popular}, and the linguistics network
\cite{ferrer01language}  
(pairs of words in English texts that appear at most one word apart). In
computer science, perhaps the first and most  
surprising result at its time was the proof that the random
graph-based models of the Internet (with their Poisson degree
distribution) were inaccurate: the Internet topology had a power-law degree
distribution \cite{faloutsos99powerlaw}. Other results followed: the web
graph \cite{barabasi00scalefree,broder00web} and the Gnutella overlay (as of
year 2000) \cite{ripeanu02mapping} are also power-law
networks. 

Another class of networks are the ``small worlds''. Two
characteristics distinguish small-world networks: first, a small average path length,
typical of random graphs (here `path' means shortest node-to-node
path); second, a large clustering coefficient that is independent of
network size. The clustering coefficient captures how many of a node's
neighbors are connected to each other. This set of characteristics is
identified in systems as diverse as social networks, in  which nodes
are people and edges are relationships; the power grid system
of western USA, in which nodes are generators, transformers,
substations, etc. and edges are transmission lines; and neural
networks, in which nodes are neurons and edges are synapses or gap
junctions \cite{watts99-book}.

\subsection{Research Questions}

Newman shows that scientific
collaboration networks in different domains (physics, biomedical research,
neuroscience, and computer science) have the characteristics  
of small worlds \cite{newman01scientific-net1,
newman01scientific-net2, newman98scientific-net}. Collaboration
networks connect scientists who have written articles together. 

Moreover, Girvan and Newman \cite{girvan02community} show
that well-defined groups (such as a research group in a specific
field) can be identified in 
(small-world) scientific collaboration networks. In parallel, a 
theoretical model for small-world networks by Watts and Strogatz
\cite{watts98collective} pictures a small world as a loosely connected
set of highly connected subgraphs.  

From here, the step is natural: since scientists tend to collaborate
on publications, they most likely use the same resources
(\textit{share} them) during their collaboration: for example, they
might use the same instruments to observe physics phenomena, or they
might analyze the same data, using perhaps the same software tools or
even a common set of computers. This means that if we connect
scientists who use the same files, we might get a small world. Even
more, we might be able to identify groups that share the same
resources. Notice that the notion of ``collaboration'' transformed
into ``resource sharing'': the social relationships do not matter
anymore, scientists who use the same resources within some time
interval may never hear of each other.

Resource sharing in a (predominantly) scientific
community is the driving force of computational Grids. If we indeed
see these 
naturally occurring sharing patterns and we find ways to exploit them (e.g., by
identifying users grouped around common sets of resources), then we
can build mechanisms 
that can tame the challenges typical 
of  large-scale, dynamic, heterogeneous, latency-affected distributed 
systems.  

The research question now become clear: 

\begin{itemize} 

\item [\textit{Q1}] \textit{Are there any patterns in the way
scientists share resources that could be exploited for designing
mechanisms?} 

\end{itemize} 

But resource sharing also exists outside scientific communities:
peer-to-peer systems or even the Web facilitate the sharing of
data. Another question arises:  

\begin{itemize} 
\item [\textit{Q2}] \textit{Are these characteristics typical of
scientific communities or are they more general?} 
\end{itemize} 

This article answers these two questions: it shows
that small-world patterns exist in diverse file-sharing communities. 

\section{The Data-Sharing Graph}
\label{sec:DSG}

To answer question \textit{Q1}, we define a new graph that captures the
virtual relationship between users who request the same data at about
the same time.  

\textit{Definition: The data-sharing graph is a
graph in which nodes are users and an edge connects two users with
similar interests in data.}

We consider one similarity criterion in this article: the number of
shared requests within a specified time interval.  

To answer question \textit{Q2}, we analyze the data-sharing graphs of
three different file-sharing communities. Section
\ref{sec:3communities} presents 
briefly these systems and the traces we used. We discover that in all
cases, for different similarity criteria, these data-sharing graphs
are small worlds. The next sections show that using the data-sharing
graph for system characterization has potential both for basic
science, because we can identify new structures emerging in real,
dynamic networks (Section \ref{sec:SWDSG}); and for system design,
because we can exploit these structures when designing data location
and delivery mechanisms (Section \ref{sec:DSG-uses}). 

\section{Three Data-Sharing Communities}
\label{sec:3communities}

We study the characteristics of the data-sharing graph corresponding
to three file-sharing communities: a
high-energy physics collaboration (Section \ref{sec:D0}), the Web as
seen from the Boeing traces (Section \ref{sec:WWW}), and the Kazaa
peer-to-peer file-sharing system seen from a large ISP in Israel (Section
\ref{sec:kazaa}). 

This section gives a brief description of each community and its
traces (duration of each trace, number of users and files requested, 
etc.) In addition, we present the file popularity and user activity
distributions for each of these traces as these have a high impact on
the characteristics of the data-sharing graph: intuitively, a 
user with high activity is likely to map onto a highly connected node
in the data sharing graph. Similarly, highly popular files are likely
to produce dense clusters. 

\begin{table}[h]
\begin{center}
\caption{Characteristics of traces analyzed.}
\begin{tabular}{|l|r|r|r|r|}
        \hline
System & Users & \multicolumn{2}{|c|}{Requests} & Duration\\
 & & All & Distinct & Traces\\
        \hline
D0 & 317& 2,757,015& 193,686& 180 days\\
Web & 60,826& 16,527,194& 4,794,439& 10 hours\\
Kazaa & 14,404 & 976,184 & 116,509 & 5 days\\
        \hline
\end{tabular}
\label{table-stats-systems}
\end{center}
\end{table}

\subsection{The D0 Experiment: a High-Energy Physics Collaboration}
\label{sec:D0}

The D0 experiment \cite{D0Experiment} is a
virtual organization comprising hundreds of physicists from more than
70 institutions in 18 
countries. Its purpose is to provide a worldwide system of shareable
computing and storage resources that can together solve the common
problem of extracting 
physics results from about a Petabyte (c.2003) of measured and
simulated data. In this system, data files are read-only and 
typical jobs analyze and produce new, processed data files. The
tracing of
system utilization is possible via a software layer (SAM
\cite{loebel-carpenter01}) that provides centralized file-based data
management. 

\begin{figure}[htpb]
\begin{center}
\includegraphics[angle=270,width=1.7in]{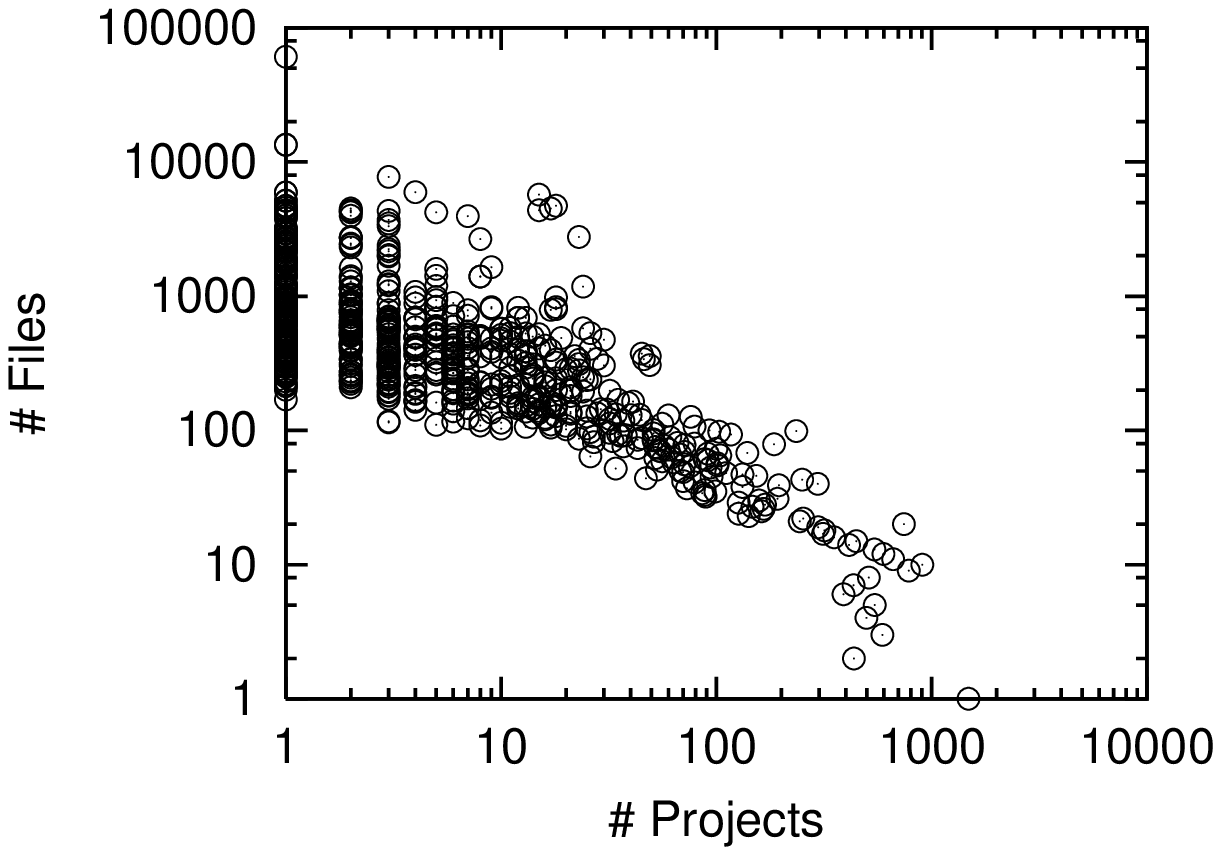}
\includegraphics[angle=270,width=1.7in]{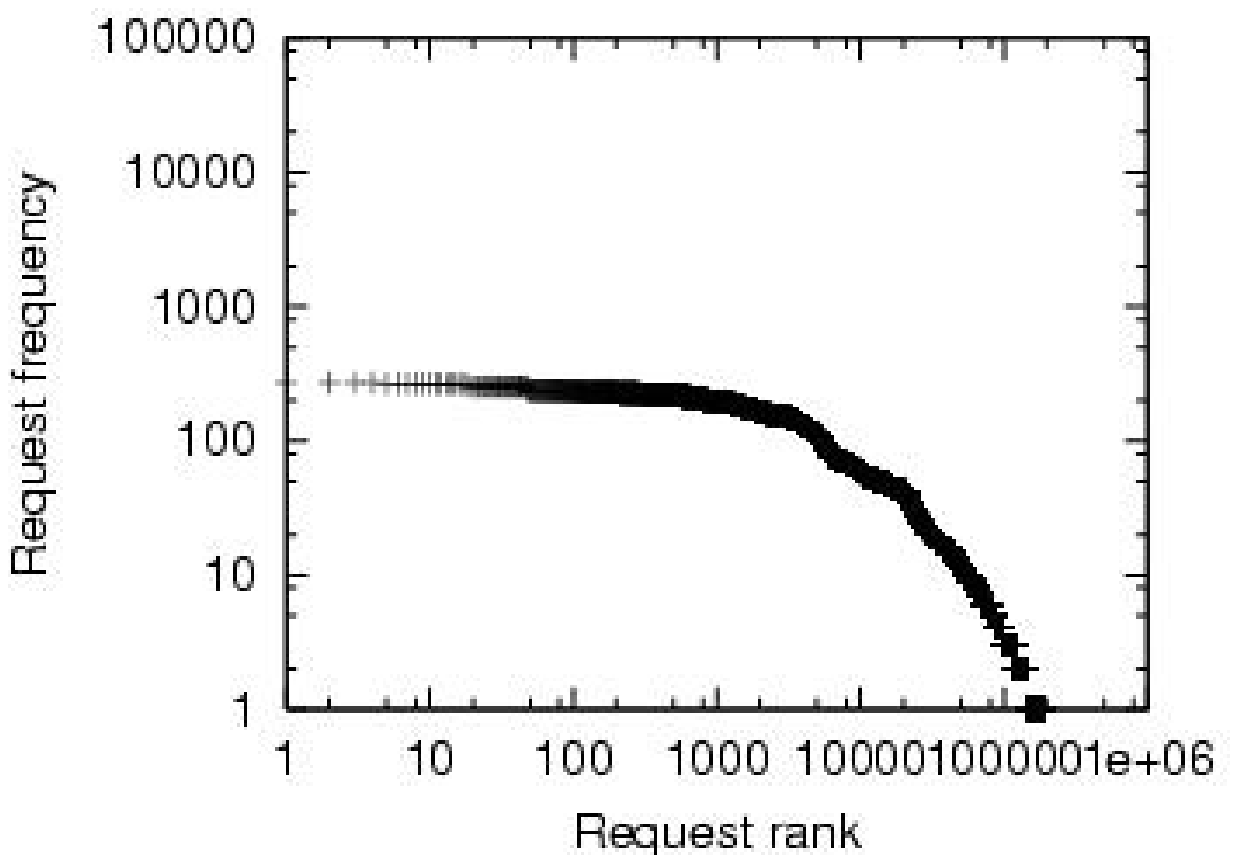}
\caption{\textit{Left:} Number of file requests per project in
D0. \textit{Right:} File popularity distribution in D0} 
\label{fig:d0-files-per-prj}
\end{center}
\end{figure}

We analyzed logs over the first six months of 2002, amounting to about
23,000 jobs submitted by more than 300 users and involving
more than 2.5 million requests for about 200,000 distinct files. A
data analysis 
job typically runs on multiple files (117 on average). Figure
\ref{fig:d0-files-per-prj} left shows the distribution of the number of
files per job.  



\begin{figure}[htpb]
\begin{center}
\includegraphics[angle=270,width=1.7in]{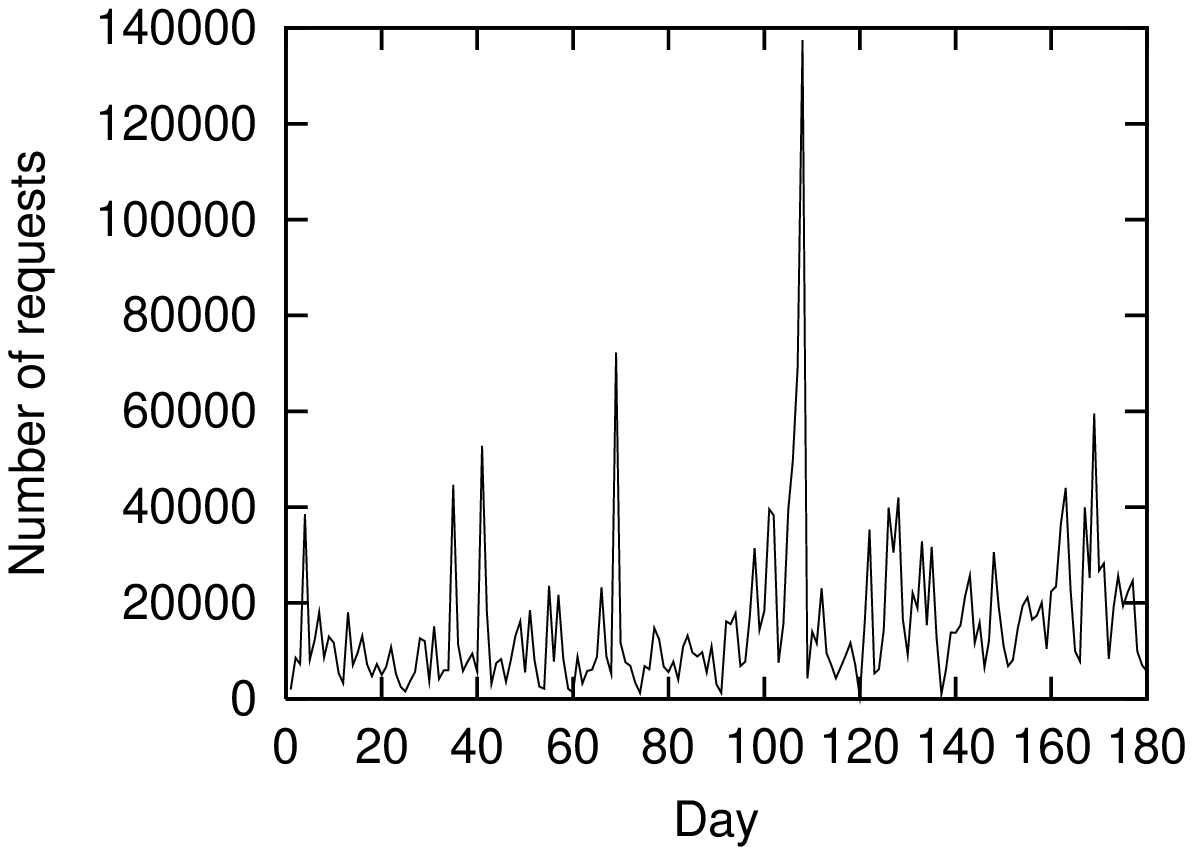}
\includegraphics[angle=270,width=1.7in]{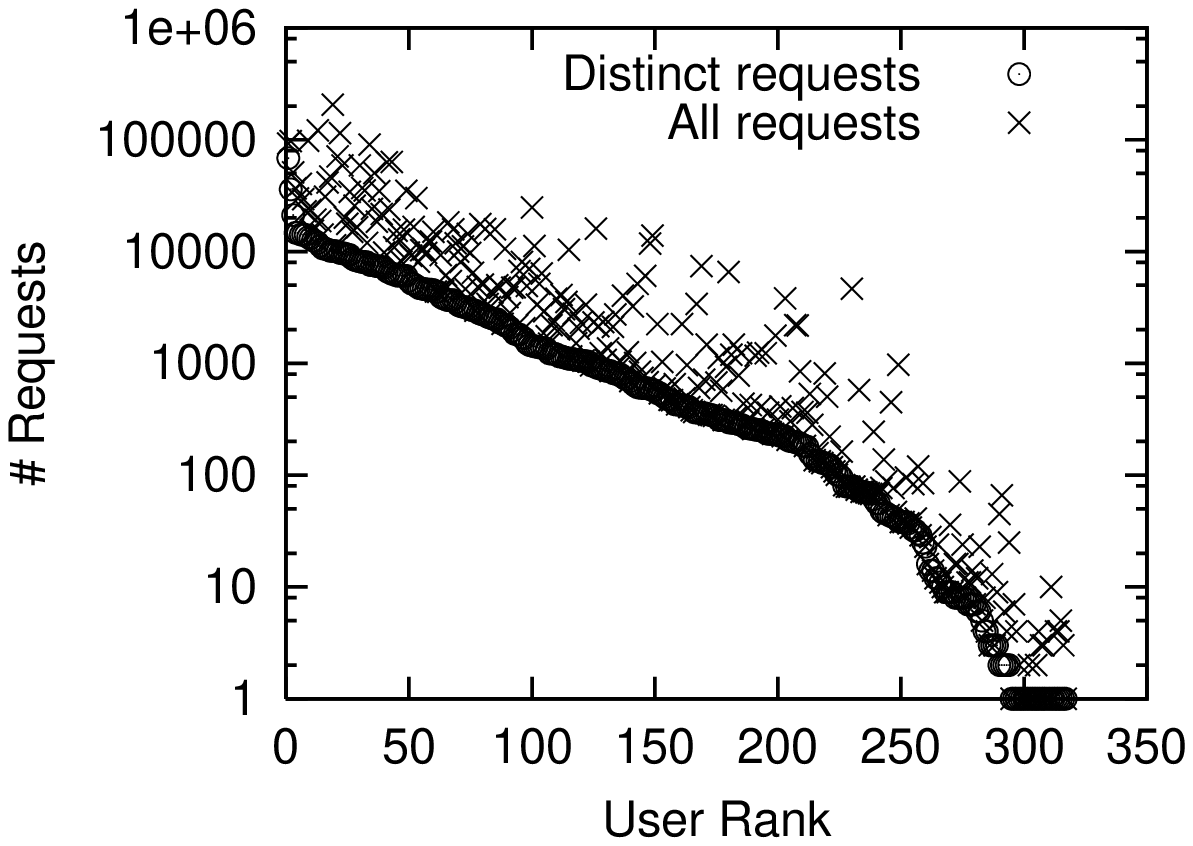}
\caption{\textit{Left:} Number of file requests per day in D0. \textit{Right:}
Number of files (total and distinct) asked by each user during the
6-month interval.} 
\label{fig:d0-reqs-per-day}
\end{center}
\end{figure}

Figure \ref{fig:d0-reqs-per-day} shows the daily activity (in number
of requests per day) and user activity (in number of requests
submitted by each user during the 6-month interval). The daily
activity is relatively constant, with a few significant
peaks---corresponding perhaps to approaching paper submission
deadlines in 
high-energy physics?. User activity is highly variable, with
scientists who scan from tens of thousands of distinct data files to
just a couple.  

In D0 file popularity does not follow the Zipf's law typical of Web
requests. (Figure \ref{fig:d0-files-per-prj}, right). The
reason we believe is that data in this scientific application is 
more uniformly interesting: a typical job swipes a significant part of
the data space (and hence file set) 
in search of particular physics events.


\subsection{The Web}
\label{sec:WWW}

\begin{figure}[htpb]
\begin{center}
\includegraphics[angle=270,width=1.7in]{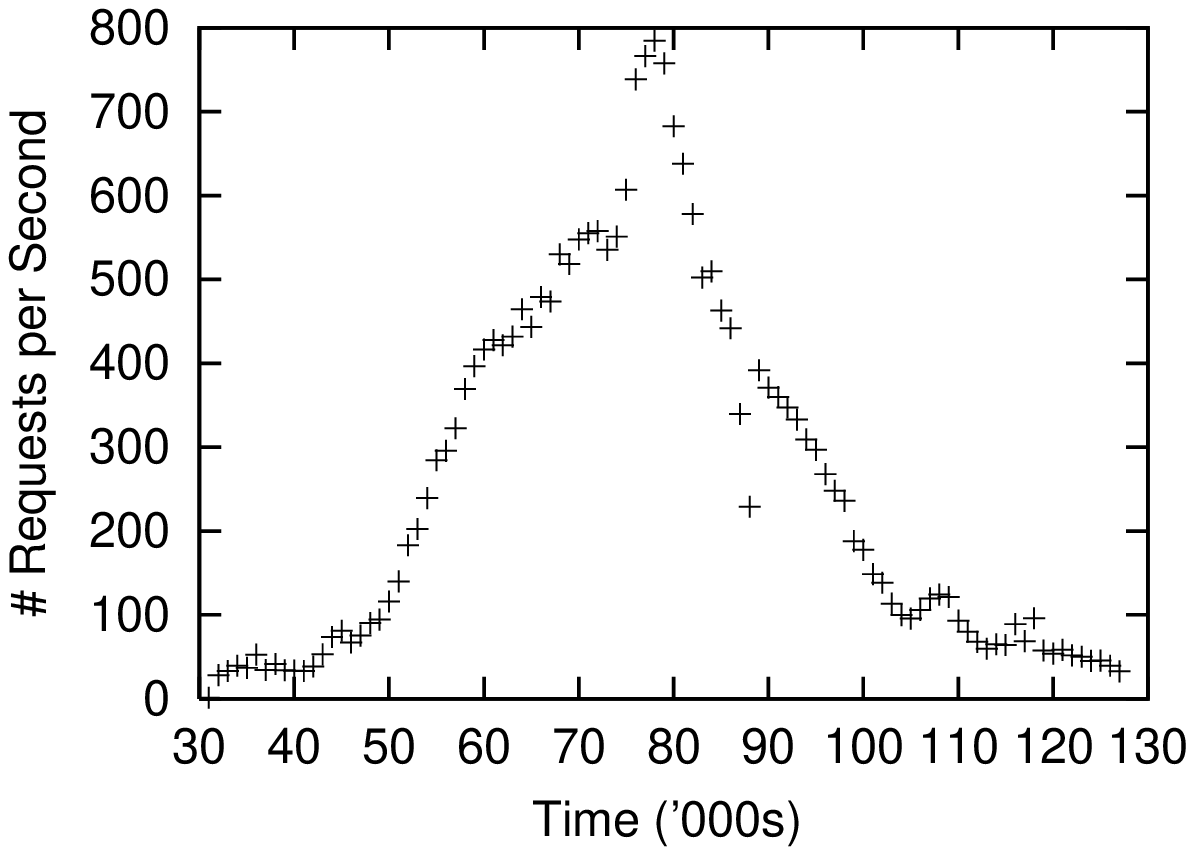}
\includegraphics[angle=270,width=1.7in]{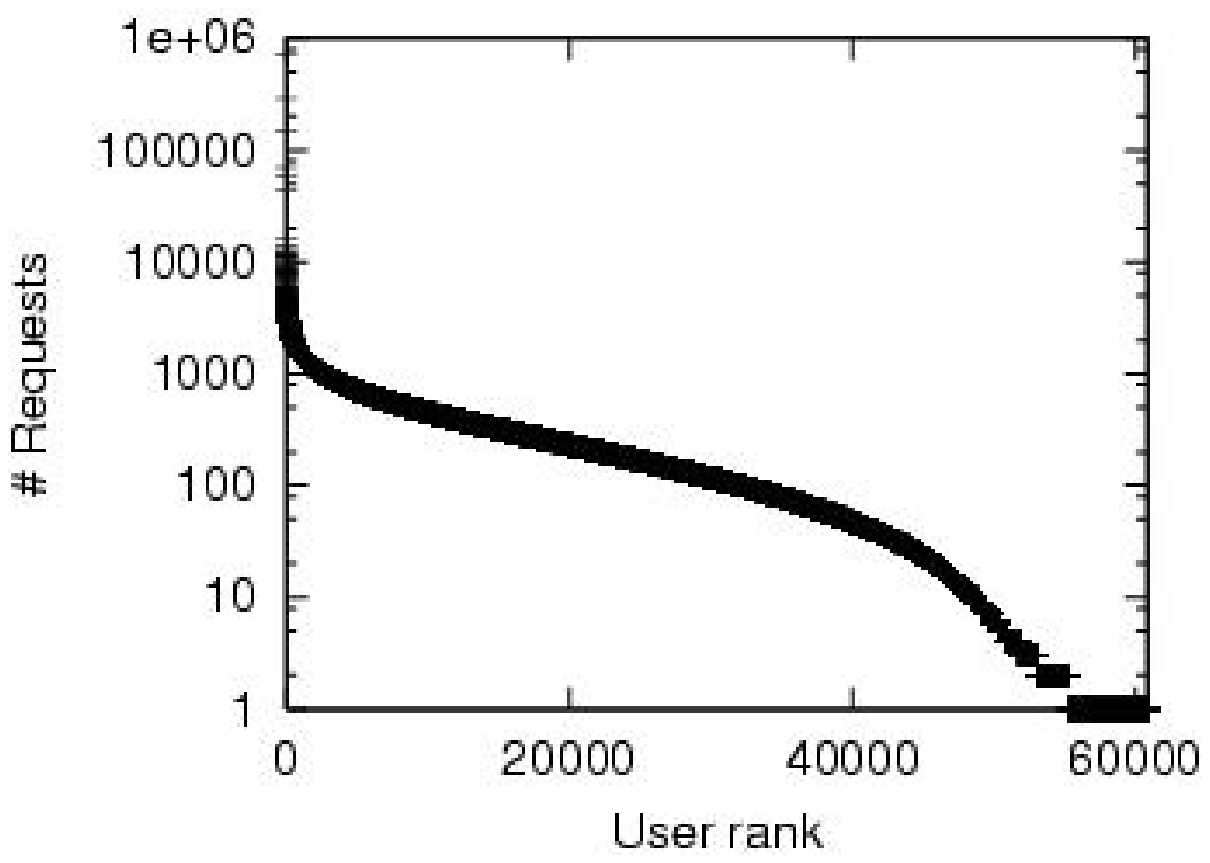}
\caption{\textit{Left:} Activity level (averaged over 15-minute
intervals). \textit{Right:} Number of requests per Web user.} 
\label{fig:web-reqs-per-user}
\end{center}
\end{figure}

We use the Boeing proxy traces \cite{boeing-traces} as a
representative sample for Web data access patterns. These traces
represent a five-day 
record from May 1999 of all HTTP requests (more than 20M requests per
day) from a large organization (Boeing) to the Web. Because traces are
anonymized and IDs are not preserved from day to day, our study
was limited to one-day intervals. However, given the intense activity
recorded (Figure \ref{fig:web-reqs-per-user} left shows the number of
requests per second), this limitation does not affect the accuracy of
our results. Here we study a representative 10-hour interval.


For the study of Web traces, we consider a user as an IP
address. During the 10-hour interval, 60,826 users sent 16.5 million
web requests, of which 4.7 million requests were distinct. It
is possible that the same IP address corresponded in fact to multiple
users (for example, for DHCP addresses or shared workstations). We do
not have any additional information to help us identify these cases or
evaluate their impact. However, given the relatively short intervals
we consider in our studies---from 2 minutes to a couple of hours---the
chances of multiple users using the same IP are small. 

\begin{figure}[htpb]
\begin{center}
\includegraphics[width=2in]{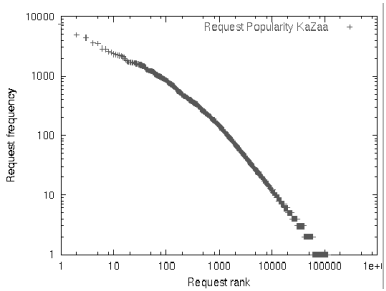}
\caption{The file popularity distributions in  Kazaa follows Zipf's
law.}
\label{fig:web-zipf}
\end{center}
\end{figure}

\subsection{The KaZaA Peer-to-Peer Network}

\label{sec:kazaa}

Kazaa is a popular peer-to-peer file-sharing system with an estimated
number of more than 4 million concurrent users as of June 2003
\cite{slyck}.   

Few details are publicly available about the Kazaa
protocol. Apparently, Kazaa nodes dynamically elect ``supernodes'' that
form an unstructured overlay network and use query flooding to locate
content. Regular nodes connect to one or more super-nodes and act as
querying clients to super-nodes. Control information, such as queries,
membership, and software version. is encrypted. Once content has been
located, data is transfered (unencrypted) directly from provider to
requester using the HTTP protocol. In order to improve
transfer speed, multiple file fragments are downloaded in parallel
from multiple providers. 

Since control information is encrypted, the only accessible traffic
information can be obtained from the download channel. As a result we
can only gather information about the files requested for download and
not about files searched for (therefore, typos are naturally
filtered). Details on how Kazaa traces were recorded as well as a
thorough 
analysis of the Kazaa traffic are presented in \cite{leibowitz03deconstructing}.


\begin{figure}[htbp]
\begin{center}
\includegraphics[angle=270,width=1.7in]{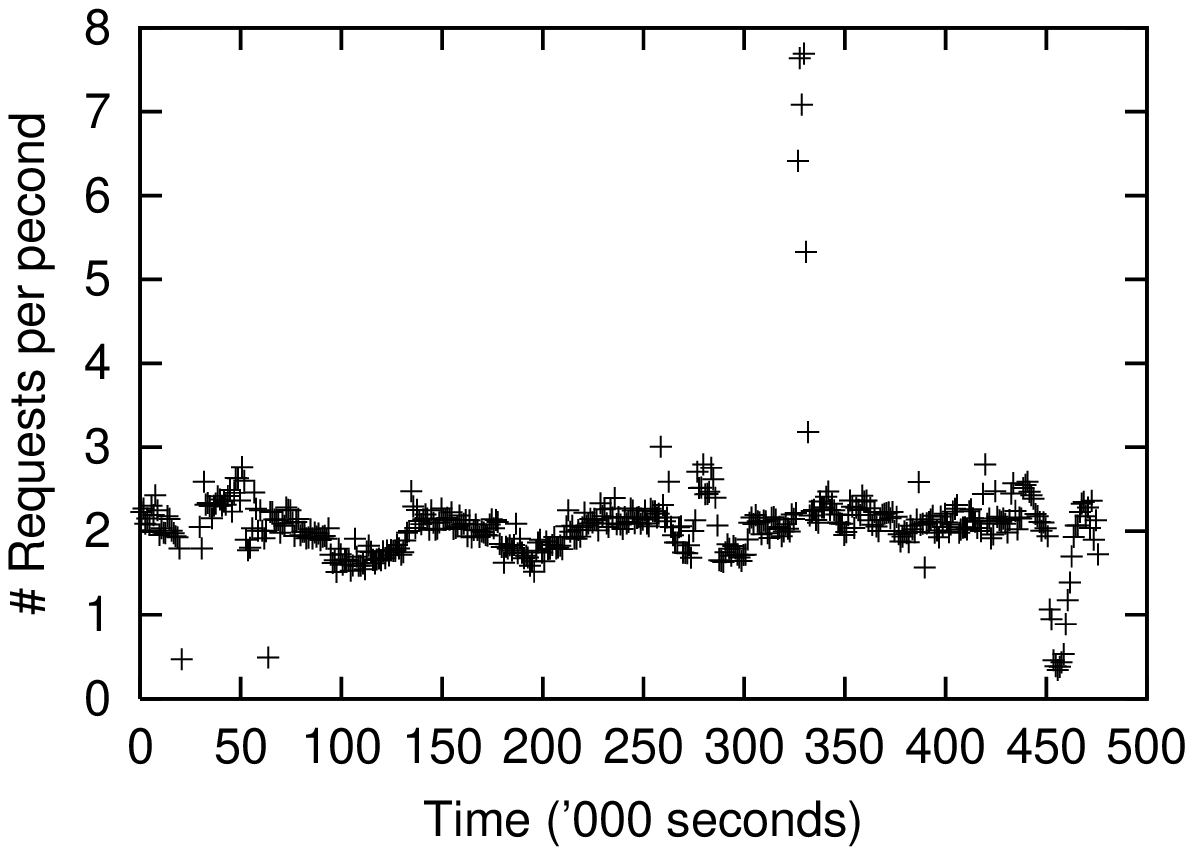}
\includegraphics[angle=270,width=1.7in]{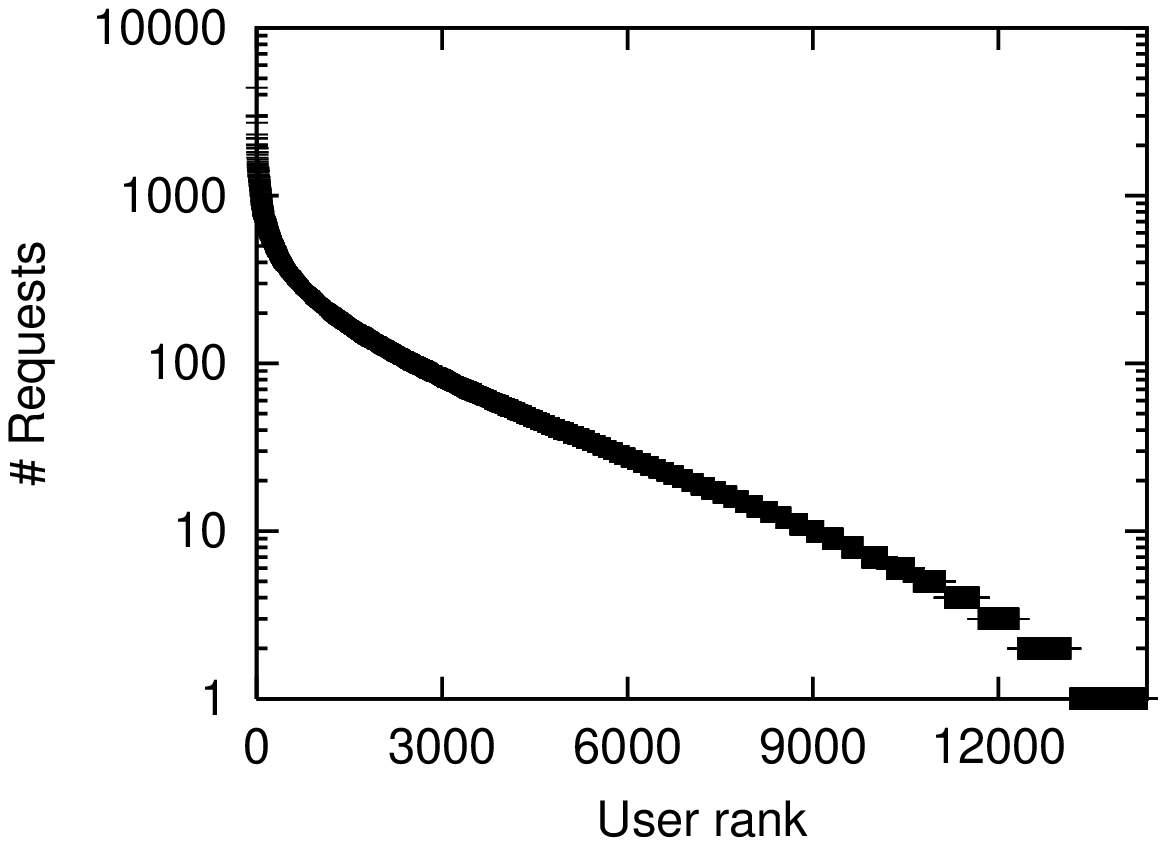}
\caption{\textit{Left:} Activity level (averaged over 100 s) in Kazaa;
\textit{Right:} Number of requests per user in KaZaa}
\label{fig:kazaa-users}
\end{center}
\end{figure}


We had access to five days of Kazaa traffic, during which 14,404 users
downloaded 976,184 files, of which 
116,509 were distinct. Users are identified based on their
(anonymized) user ID that appears in the HTTP 
download request. The user population is formed of Kazaa users who are
clients of the ISP: similar to the
Boeing traces, these traces give information about only a limited set
of Kazaa users.  

\section{Small-World Data-Sharing Graphs}

\label{sec:SWDSG}

Data-sharing graphs are built using the definition in
Section \ref{sec:DSG}: users are nodes in the graph and two
users are connected if they have similar interests in data during some
interval. For the rest of this paper we consider one class of
similarity criteria: we say that two users have similar data interests if
the size of the intersection of their request sets is larger than some
threshold. This section presents the properties of data-sharing graphs
for the three communities introduced previously. 

The similarity criterion has two degrees of freedom: the
length of the time interval and the threshold on the number of common
requests. Section \ref{sec:wd} studies
the dependence between these parameters for each of the three
data-sharing communities.

Sections \ref{sec:dd} and \ref{sec:sw} present the properties of
the data-sharing graphs. We shall see that not all
data-sharing graphs are power law. However, they all
exhibit small-world characteristics, a result that we support with
more rigorous analysis in Section \ref{sec:PrefNet}.

\subsection{Distribution of Weights}
\label{sec:wd}

\begin{figure}[htpb]
\begin{center}
\includegraphics[angle=270,width=1.6in]{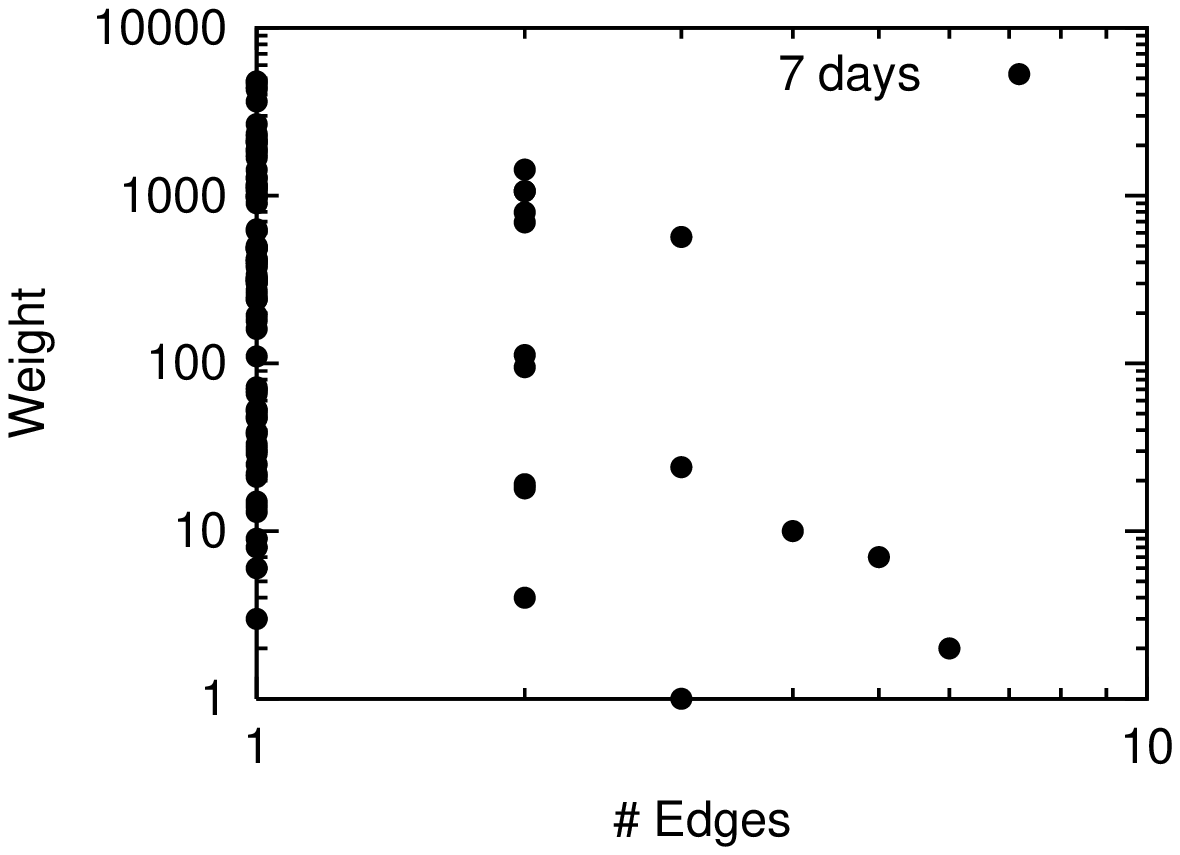}
\includegraphics[angle=270,width=1.6in]{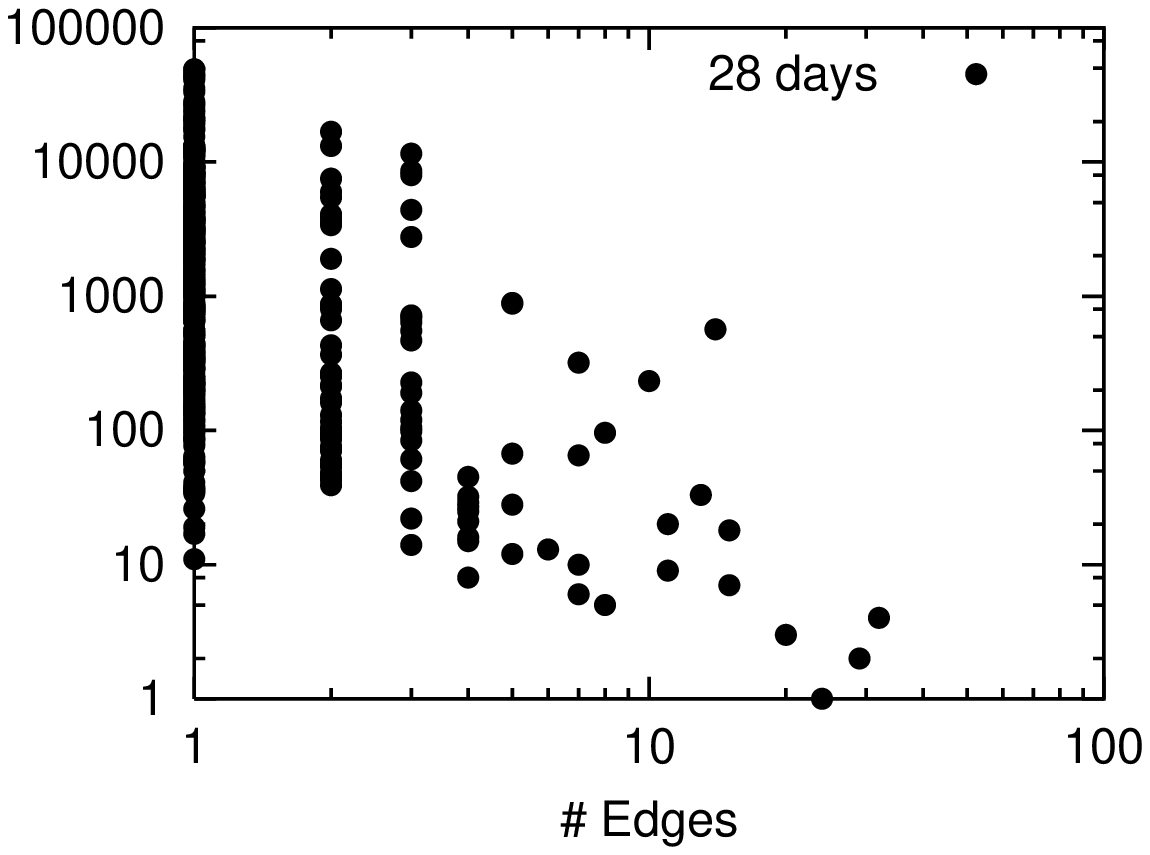}
\caption{The distribution of weights in D0 data-sharing graphs for
different intervals during the same period.}
\label{fig:d0-weightdistrib}
\end{center}
\end{figure}

\begin{figure}[htpb]
\begin{center}
\includegraphics[angle=270,width=1.7in]{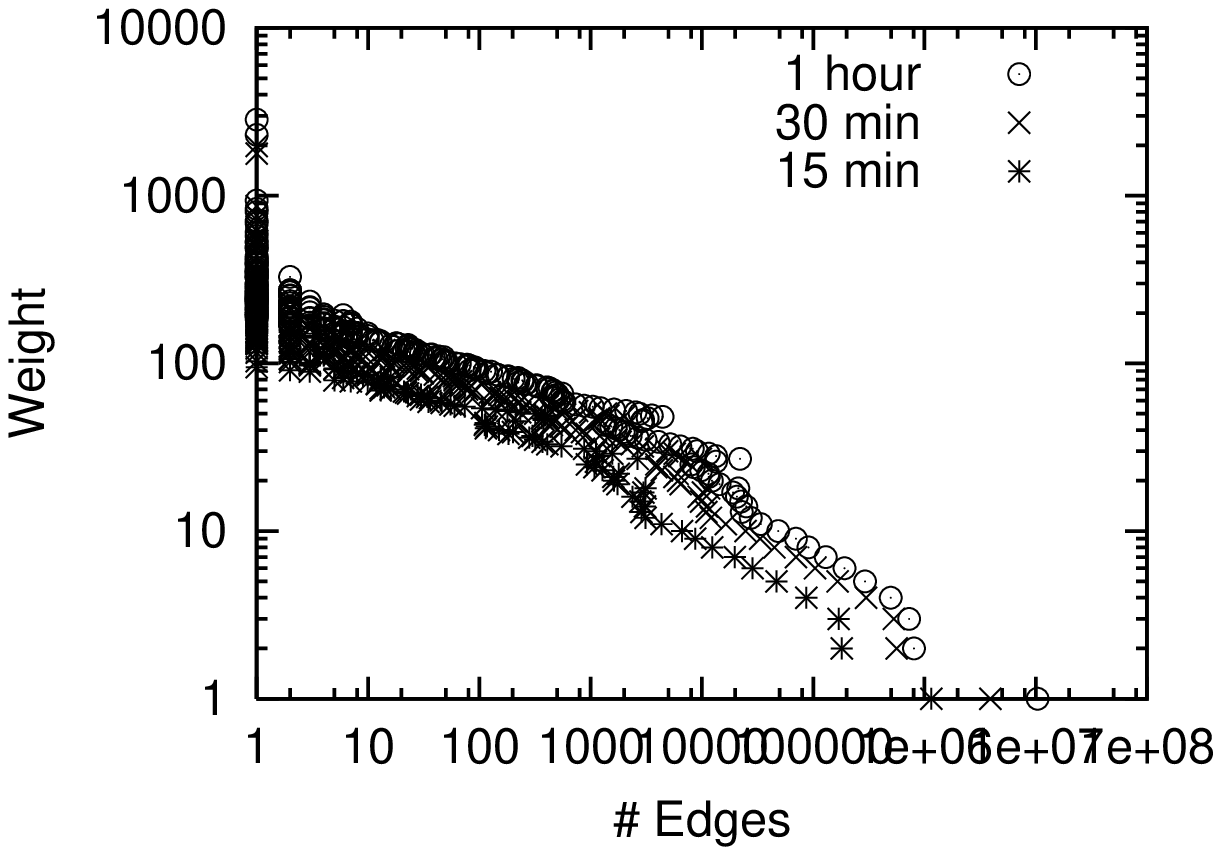}
\includegraphics[angle=270,width=1.7in]{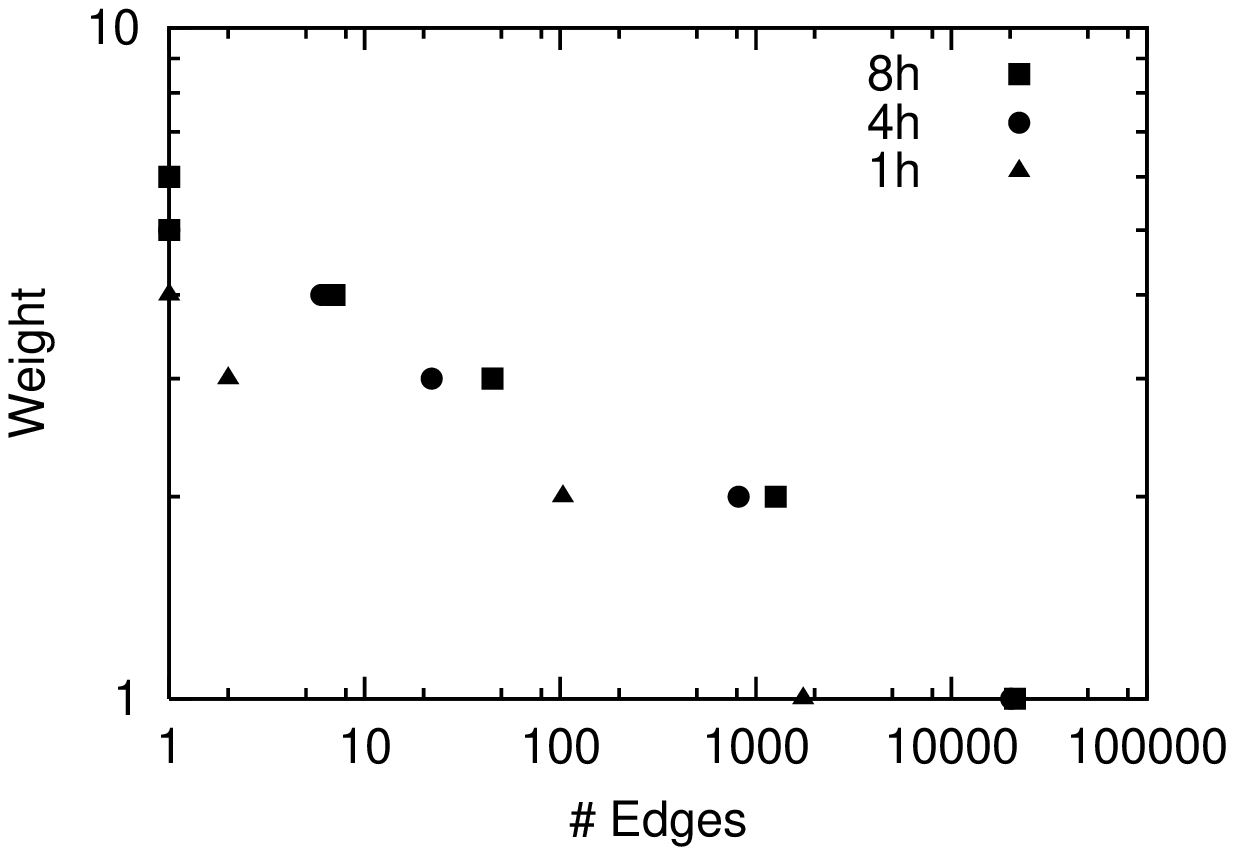}
\caption{The distribution of weights in Kazaa (left) and Web (right)
data-sharing graphs for different time intervals.}
\label{fig:web-weightdistrib}
\end{center}
\end{figure}

We can think of data-sharing graphs as weighted graphs: two
users are connected by an edge labeled with the number of shared 
requests during a specified time period. Remove 0-weight edges, as
well as isolated nodes 
(those that have no edges). We obtain a weighted data-sharing
graph (Figures \ref{fig:d0-weightdistrib} and
\ref{fig:web-weightdistrib}). The distribution of weights highlights
differences among the 
sharing communities: the sharing in D0 is significantly more
pronounced than in Kazaa, with weights in the order of hundreds or
thousands in D0 compared to 5 in Kazaa. 


\subsection{Degree Distribution}
\label{sec:dd}

\begin{figure}[htbp]
\begin{center}
\includegraphics[angle=270,width=1.7in]{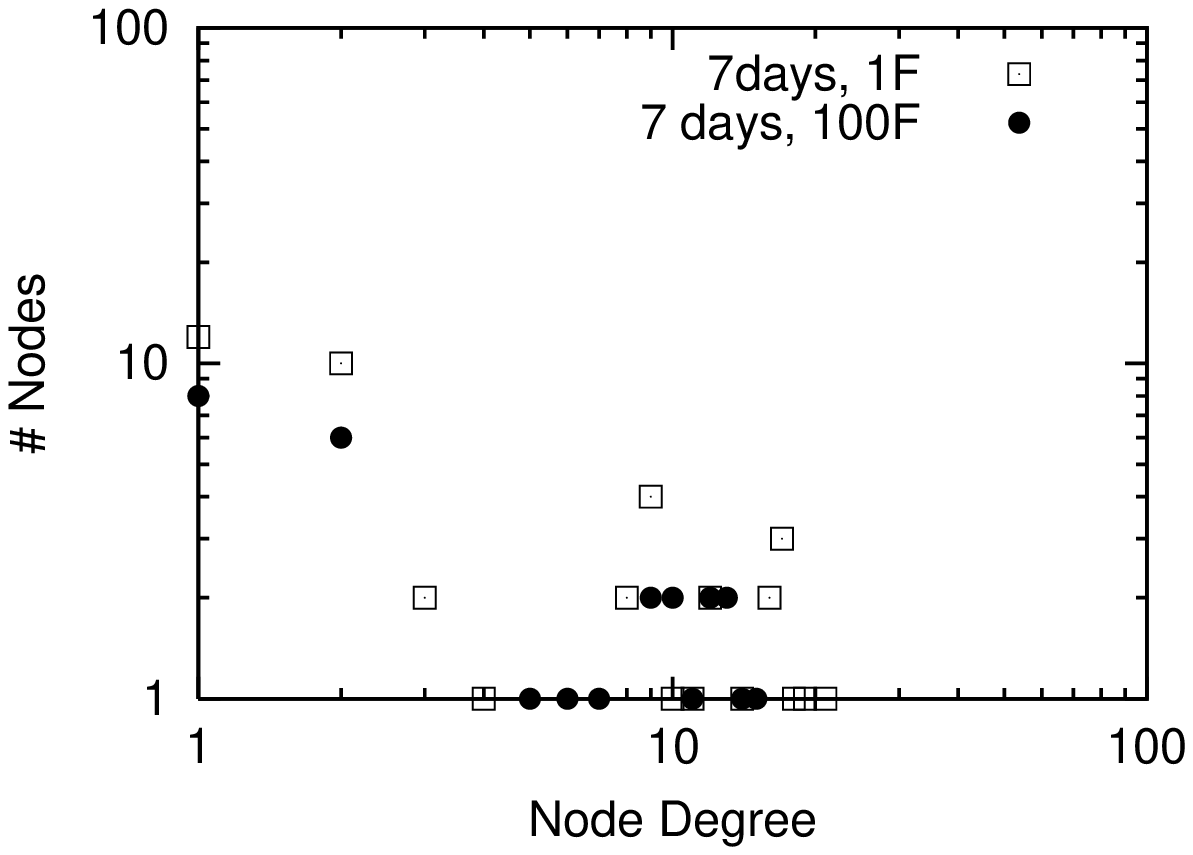}
\includegraphics[angle=270,width=1.7in]{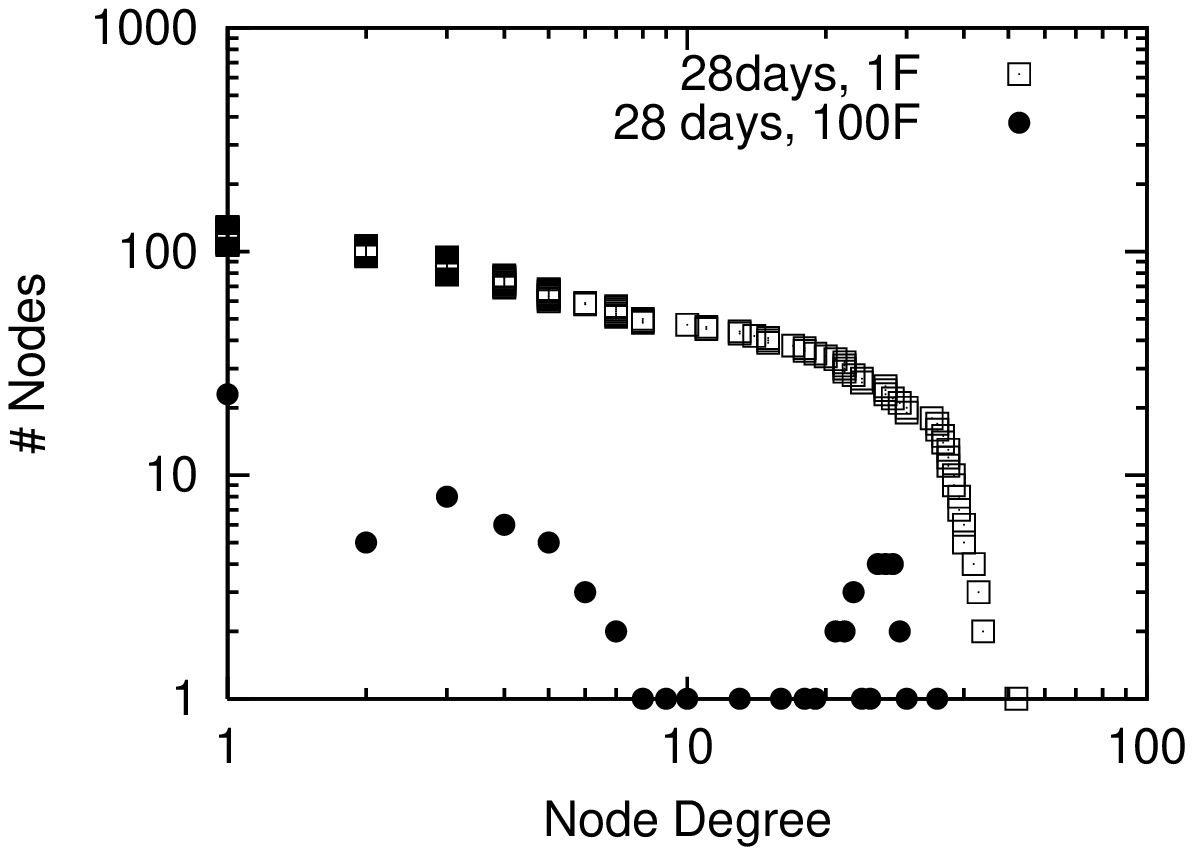}
\caption{Degree distribution for D0 data-sharing graphs.}
\label{fig:d0-dd}
\end{center}
\end{figure}

\begin{figure}[htbp]
\begin{center}
\includegraphics[angle=270,width=1.7in]{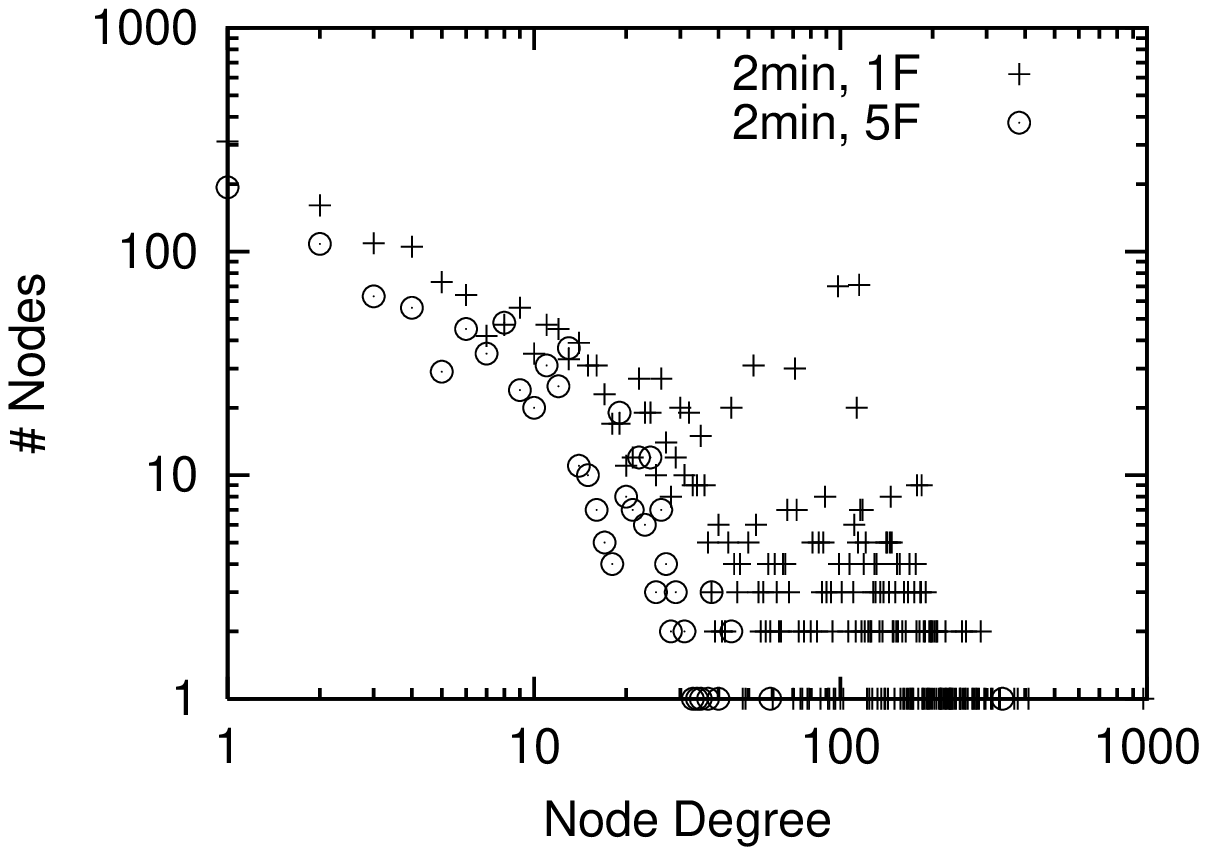}
\includegraphics[angle=270,width=1.7in]{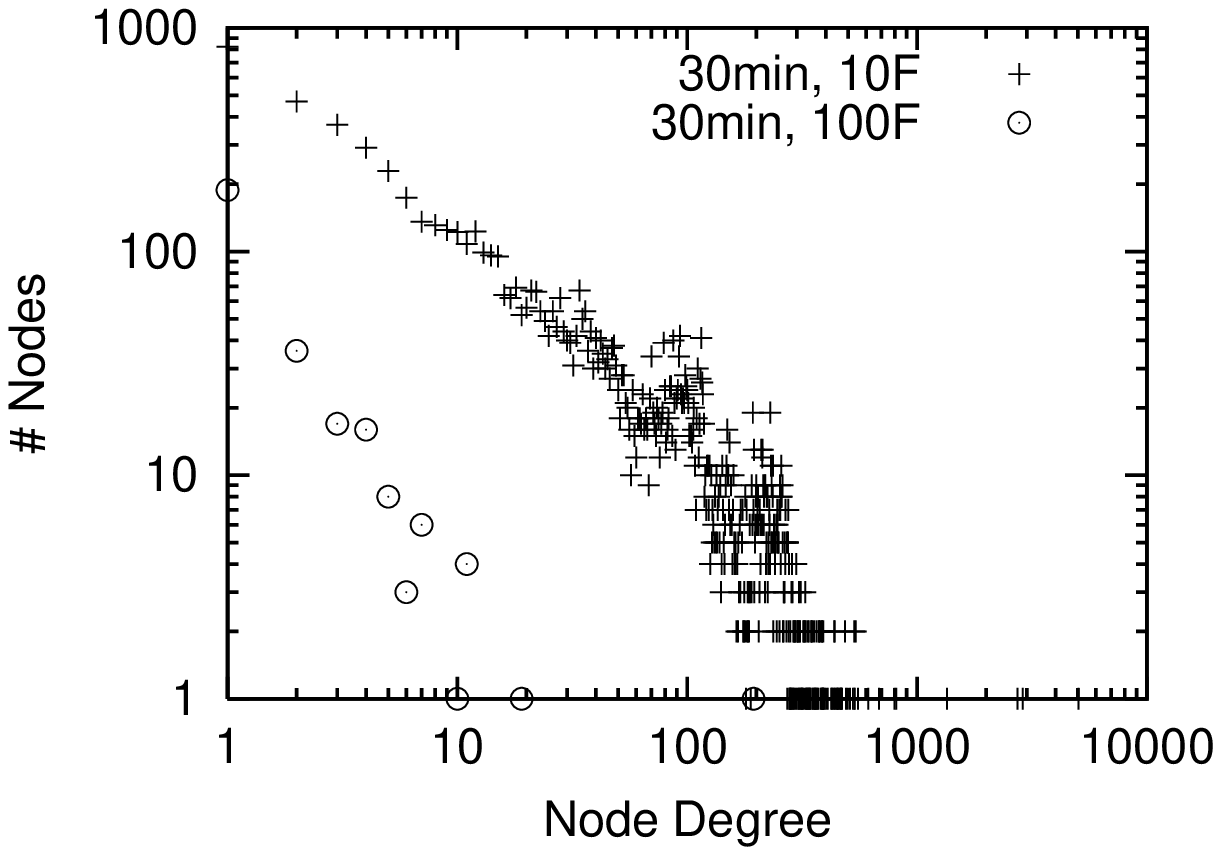}
\caption{Degree distribution for Web data-sharing graphs}
\label{fig:web-dd}
\end{center}
\end{figure}

\begin{figure}[htbp]
\begin{center}
\includegraphics[angle=270,width=1.7in]{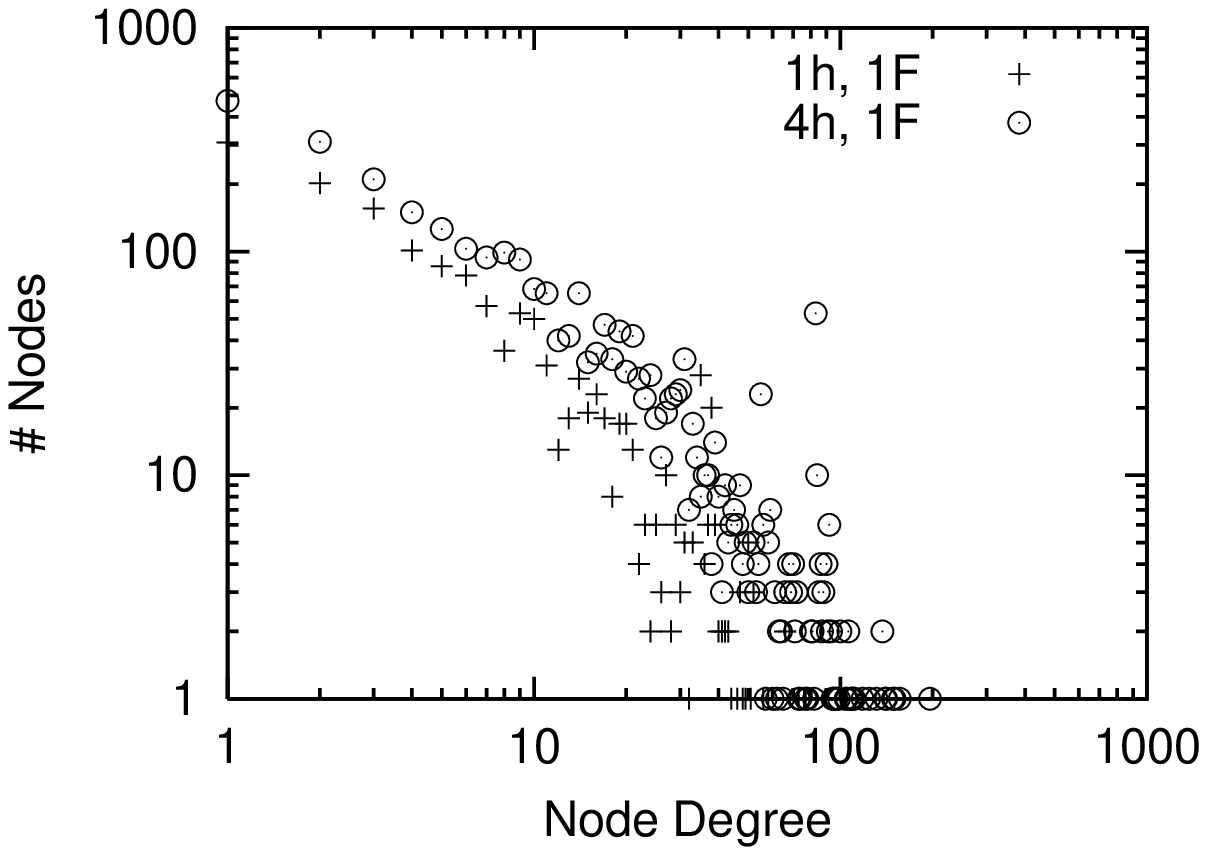}
\includegraphics[angle=270,width=1.7in]{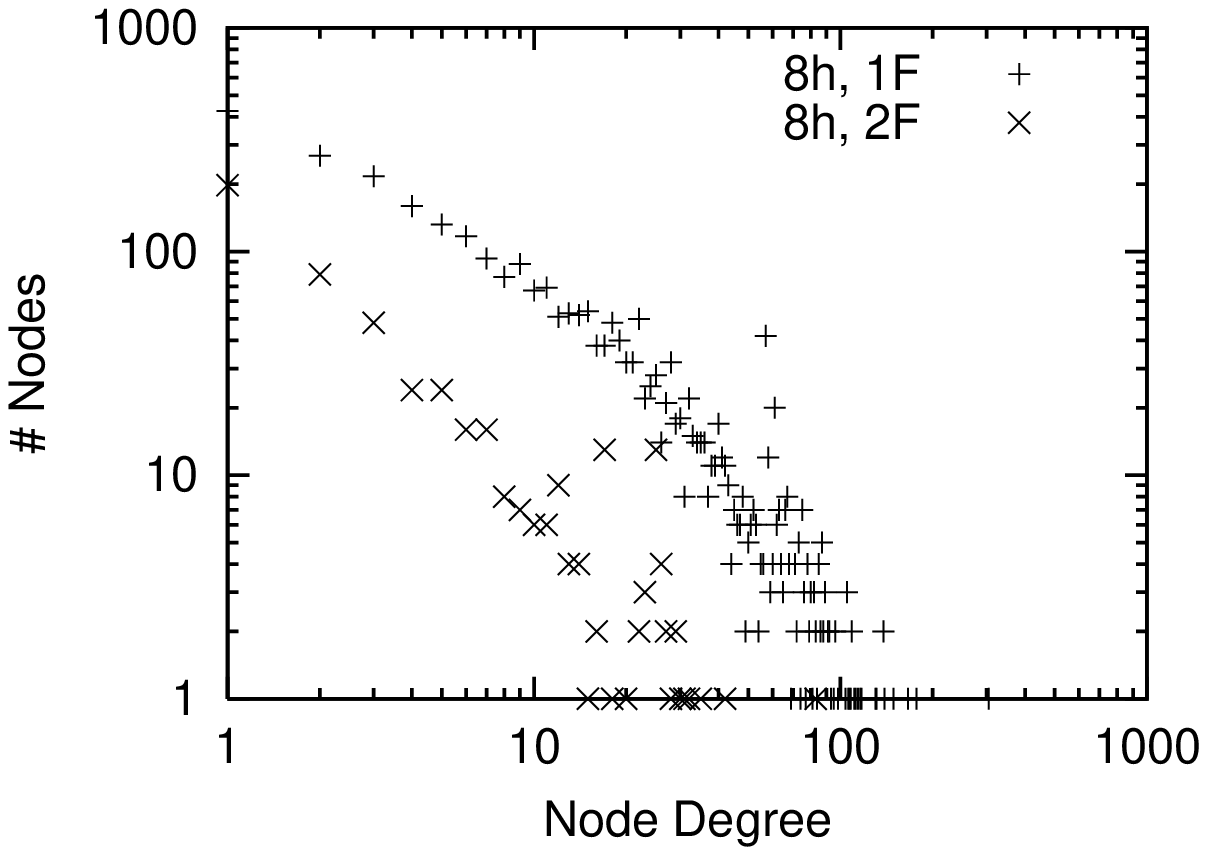}
\caption{Degree distribution for Kazaa data-sharing graphs}
\label{fig:kazaa-dd}
\end{center}
\end{figure}

The node degree distribution of the data-sharing graph is particularly
interesting for designing distributed 
applications. Figures \ref{fig:d0-dd}, \ref{fig:web-dd}, and
\ref{fig:kazaa-dd} present the degree 
distributions for the three systems: note that the Kazaa data-sharing
graph is the closest to a power-law, 
while D0 graphs clearly are not power-law.

\subsection{Small-World Characteristics: Clustering Coefficient and Average Path Length}

\label{sec:sw}

\begin{sidewaystable}
\begin{center}
\caption{Properties of data-sharing graphs for the three communities
studied. $CC_1$ is the measured Watts-Strogatz clustering coefficient
(Eq. \ref{eq:CC-Watts}), $CC_2$ is the 
measured clustering coefficient defined in Eq. \ref{eq:CC2}; $CC_r$ is the
Watts-Strogatz clustering coefficient of random graphs
(Eq. \ref{eq:CC-random}); $l$ is the measured average path length and $l_r$
is the average path length of random graphs (Eq. \ref{eq:AvgPL-random})}

\begin{tabular}{|c|r|r|r|r|r|r|r|r|r|r|r|r|}
        \hline
& Time & Files in & & & \#Connected& \multicolumn{5}{|c|}{Largest
connected component} & \multicolumn{2}{|c|}{Random Graph}\\
System & Interval& Common & \# Nodes & \# Edges & Components&  \# Nodes & \# Edges &
$CC_1$ & $CC_2$ & $l$ & $CC_r$ & $l_r$\\
        \hline
D0 & 7 days & 1  & 46 & 153 & 5  & 35& 142 & 0.741& 0.648&
2.114 & 0.238 & 2.538 \\
D0 & 7 days & 100 &   26  &    95  &    3  &     20 &     88     &
0.773 & 0.743&  1.65 & 0.463 & 2.021 \\
D0 & 7 days & 1000  & 14  &    43   &   1   &    14  &    43  &
0.834 & 0.652 & 1.5 & 0.472 & 2.351\\

D0 & 28 days &  1 & 129 & 777  &  9  & 107   &  757   &  0.716 & 0.641
& 2.476 & 0.133 & 2.388 \\
D0 & 28 days & 100  & 84  &  441 &  4  &  78  & 438  & 0.706 & 0.763
&
2.769 &  0.145 & 2.524\\
D0 & 28 days &  1000  & 49 & 235 & 6 & 35 & 226 &  0.838 & 0.808&
1.628 & 0.379 & 1.906 \\

        \hline

Web & 120 s &  1 & 2076  &  47610 &  100  &   1805  &  47256 &
0.786 & 0.634& 2.666 & 0.029 & 2.296\\

Web & 1800 s & 1 & 6137 & 1866338 & 39 & 6049 & 1866271 & 0.808
& & 2.056 &  0.102  & 1.519 \\


Web & 1800 s & 100 & 143 & 196 & 20 & 102&172
& 0.720 & 0.130 & 3.6 & 0.033 &   8.851\\




        \hline

Kazaa & 1 hour & 1 &  809 & 1937 & 97 & 548 &
1690 & 0.740 & 0.593 & 5.629 & 0.011  &  5.599\\
Kazaa & 8 hours & 1 & 3608 & 31252  & 79 & 3403 &
30555 & 0.652  & 0.473 & 3.611 & 0.005        &  3.705 \\
Kazaa & 8 hours & 3 &   111   &  111   &  24    &  56
& 78 &  0.442 & 0.178 & 3.160 & 0.050  &
12.148 \\
        \hline

\end{tabular}
\label{table:all}
\end{center}
\end{sidewaystable}

We wanted to test our intuition that, similar to scientific
collaboration networks, we find small-world patterns at the resource
sharing level. We consider
the Watts-Strogatz definition \cite{watts98collective}: a graph $G(V,E)$ is a
small world if it has small average path length and large
clustering coefficient, much larger than that of a random graph with the
same number of nodes and edges. 

The clustering coefficient is a measure of how well connected a node's
neighbors are with each other. According to one commonly used formula
for computing the clustering coefficient of 
a graph (Eq. \ref{eq:CC-Watts}), the clustering coefficient of
a node is the ratio of the number of existing edges and the
maximum number of possible edges  connecting its neighbors. The average
over all $|V|$ nodes gives the clustering coefficient of a graph
(Eq. \ref{eq:CC-avg}).  

\begin{equation}
CC_u = \frac{\mbox{\# edges between \textit{u}'s neighbors}}{\mbox{Maximum
\# edges between \textit{u}'s neighbors}}
\label{eq:CC-Watts}
\end{equation}

\begin{equation}
CC_1 = \frac{1}{|V|}\sum_u CC_u 
\label{eq:CC-avg}
\end{equation}

Another definition 
(Eq. \ref{eq:CC2}) directly calculates the clustering coefficient of a
graph as a ratio of the number of triangles and the number of
triples of connected nodes, where connected triples of vertices are
trios of nodes in which at least one is connected to the other two. 

\begin{equation}
CC_2 = \frac{3 \times \mbox{Number of triangles on the graph}}{\mbox{Number of
connected triples of vertices}}
\label{eq:CC2}
\end{equation}

The two definitions of the clustering coefficient simply reverse the
operations---one takes the mean of the ratios, while the other takes
the ratio of the means. The former definition tends therefore to weight
the low-degree vertices more heavily, since they have a small
denominator in Eq. \ref{eq:CC-Watts}. 

According to the definition of clustering from Eq. \ref{eq:CC-Watts},
the clustering coefficient of a random graph is: 

\begin{equation}
CC_r = \frac{2 \times |E|}{|V| \times (|V| -1)}
\label{eq:CC-random}
\end{equation}

The average path length of a graph is the average of all
distances. For large graphs, measuring all-pair distances is
computationally expensive, so an accepted procedure
\cite{watts99-book} is to measure it over a random sample of
nodes. The average path length for the larger Web data-sharing graphs 
in Table \ref{table:all} was approximated using a random sample of 5\%
of the graph nodes. The average path length of a random graph is given
by Eq. \ref{eq:AvgPL-random}.  

\begin{equation}
l_r = \frac{log(|V|)}{log(|E|/|V|)}
\label{eq:AvgPL-random}
\end{equation}

\begin{figure}[htpb]
\begin{center}
\includegraphics[angle=270,width=1.7in]{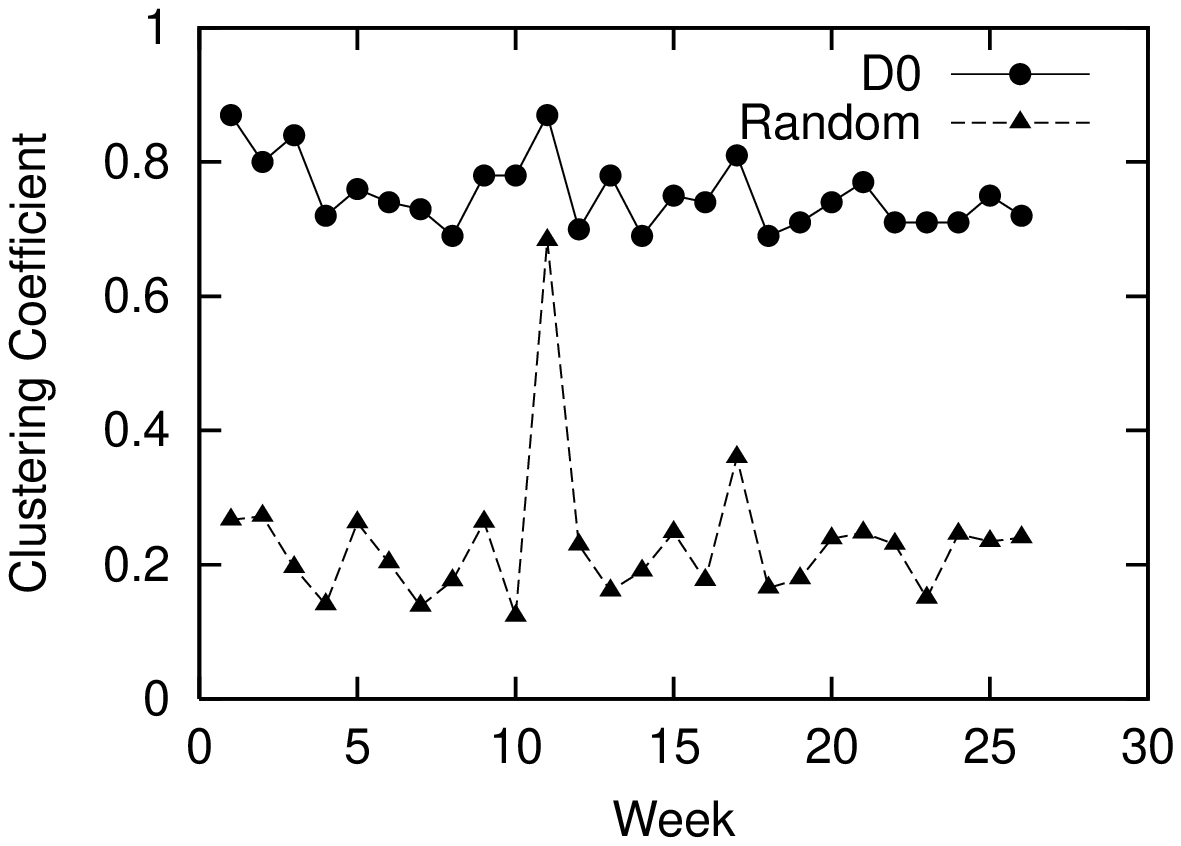}
\includegraphics[angle=270,width=1.7in]{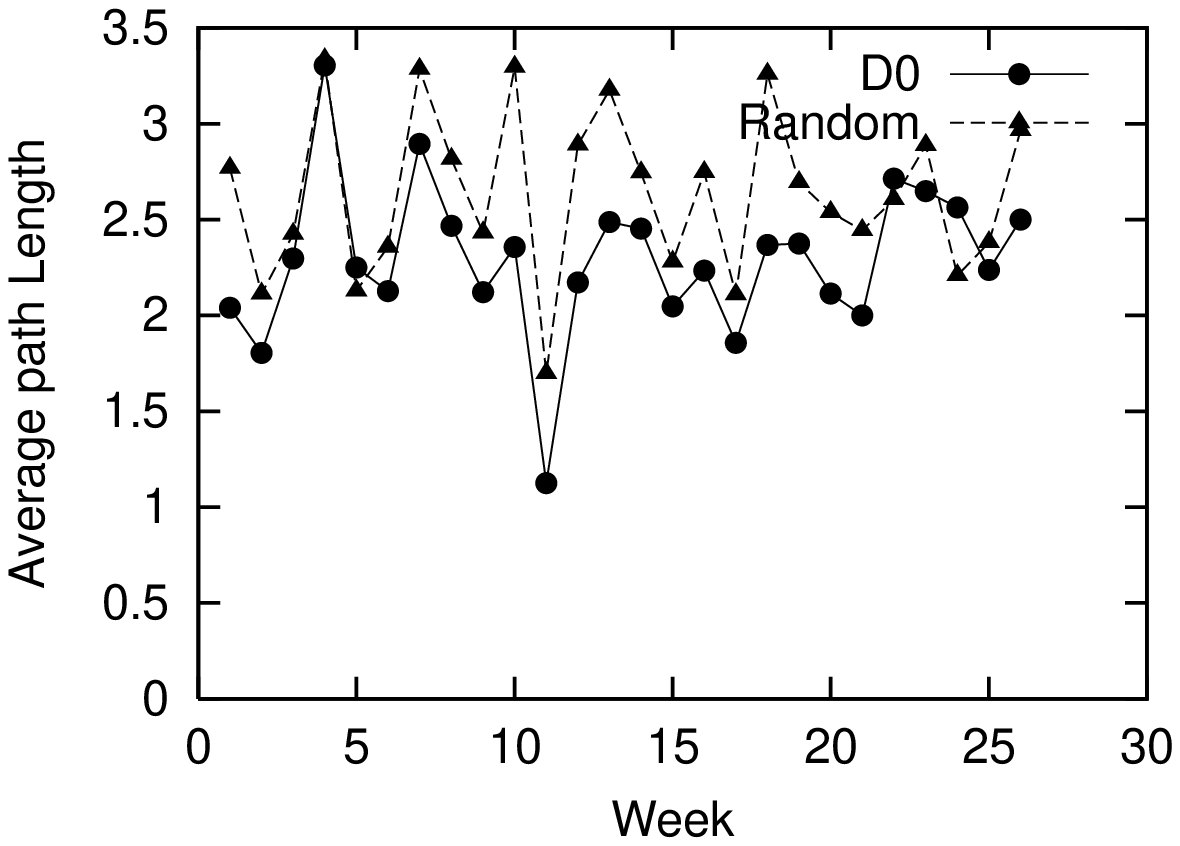}
\caption{Clustering coefficients (left) and average path lengths (right) of
D0 data-sharing graphs and random graphs of same size. Similarity
criterion: 1 shared file during a 7-day interval.}
\label{fig:d0-cc}
\end{center}
\end{figure}

\begin{figure}[htpb]
\begin{center}
\includegraphics[angle=270,width=1.7in]{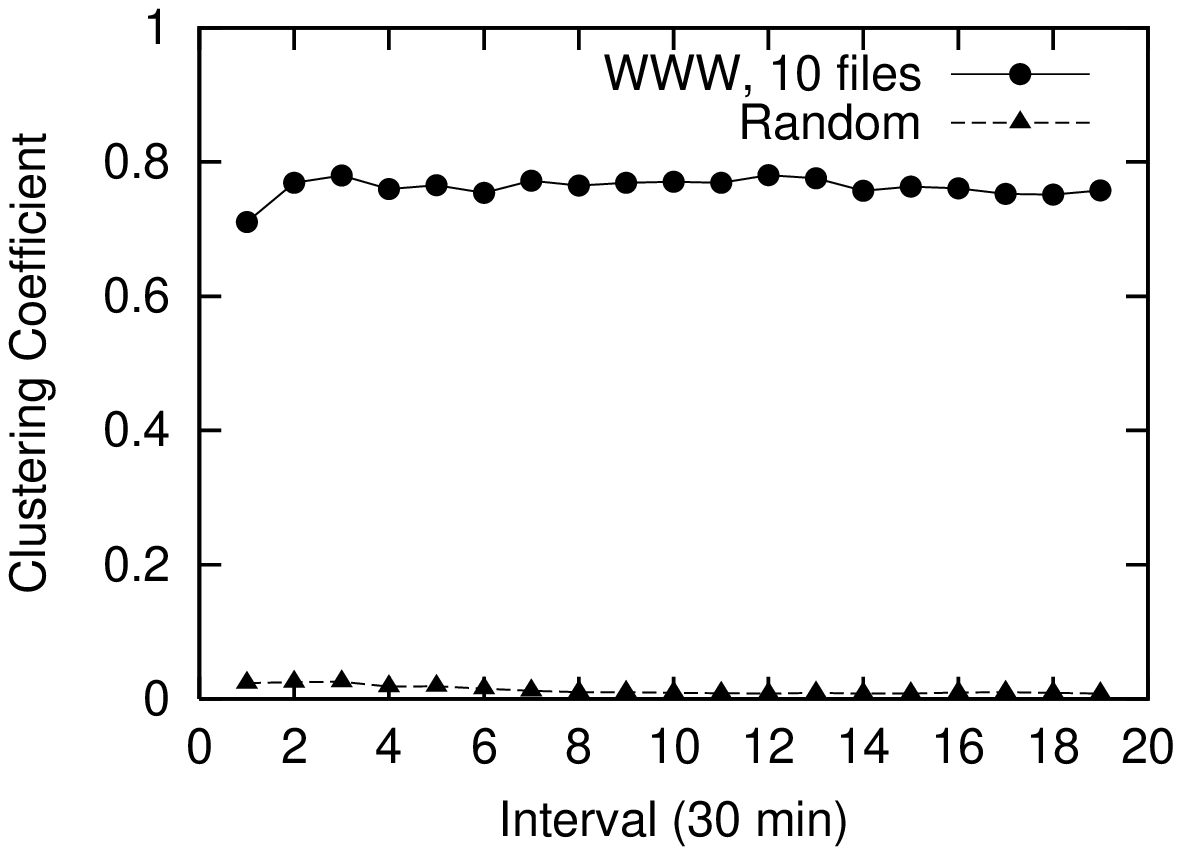}
\includegraphics[angle=270,width=1.7in]{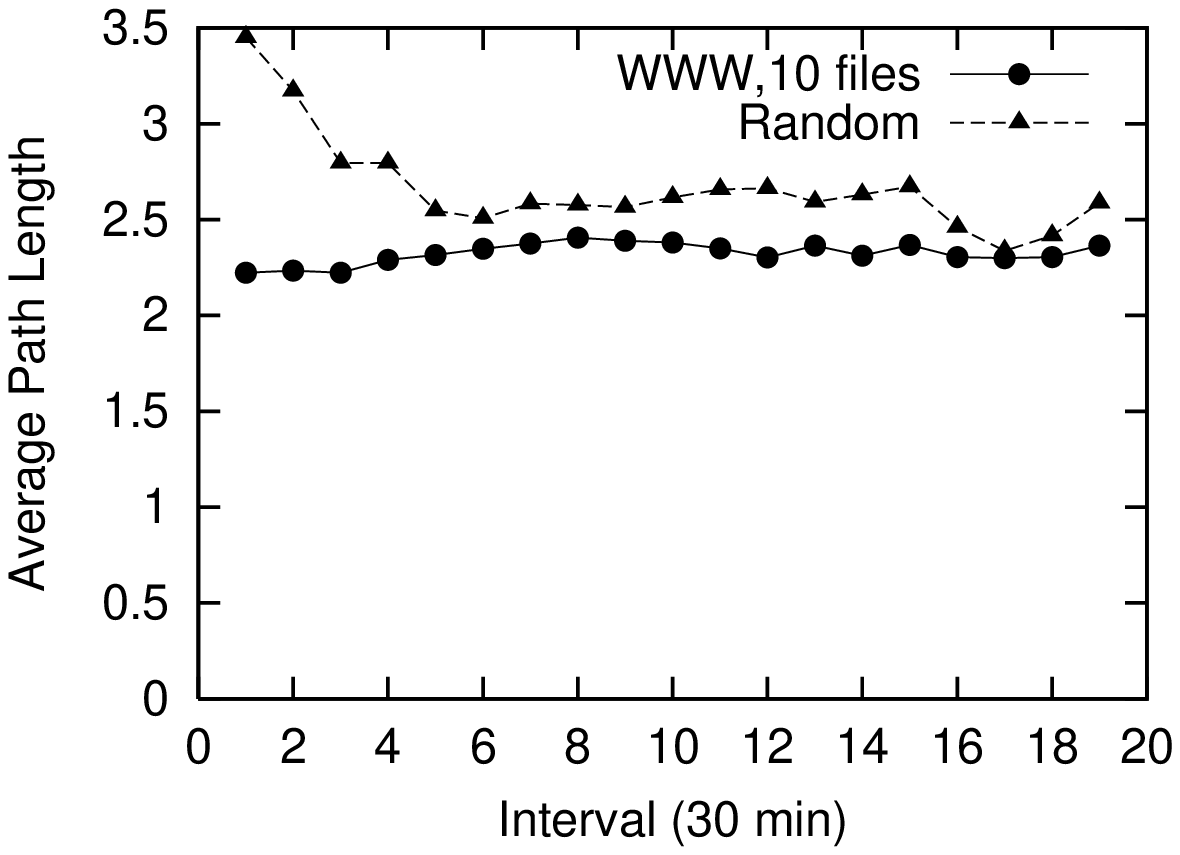}
\includegraphics[angle=270,width=1.7in]{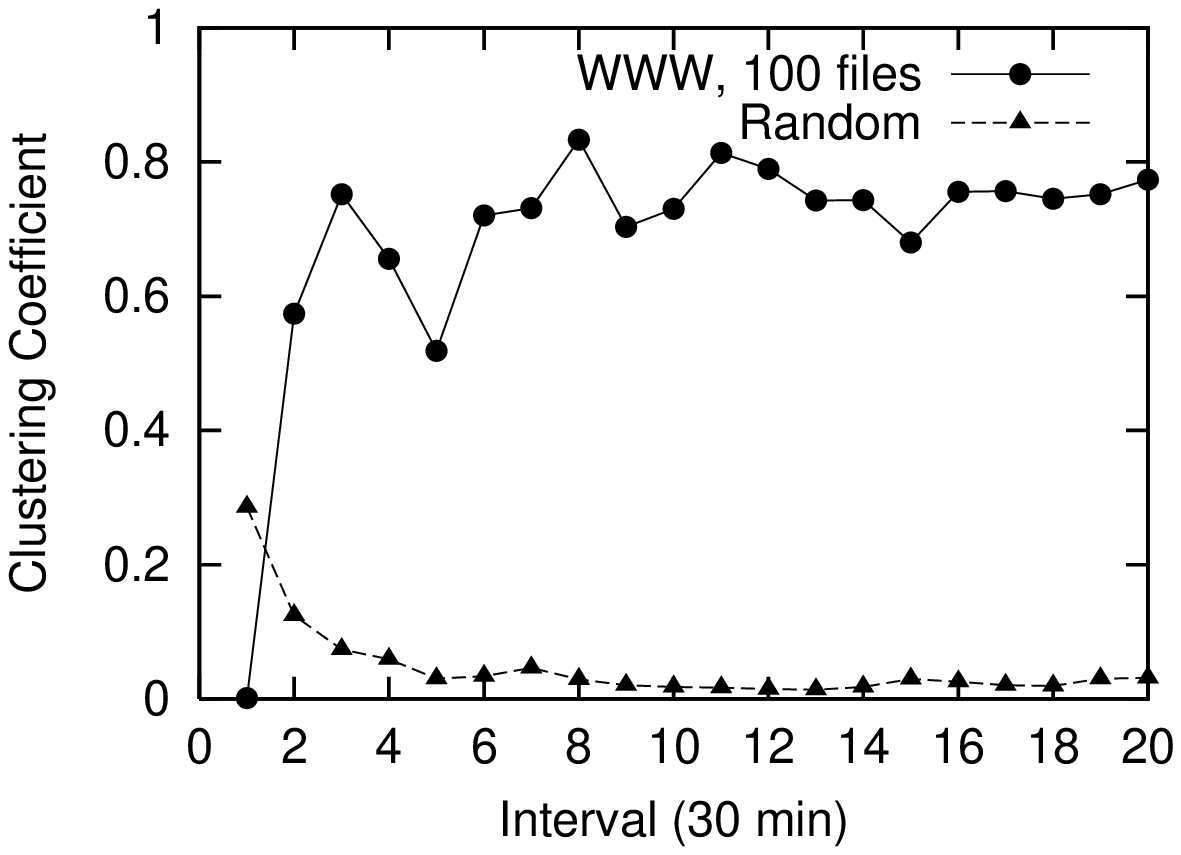}
\includegraphics[angle=270,width=1.7in]{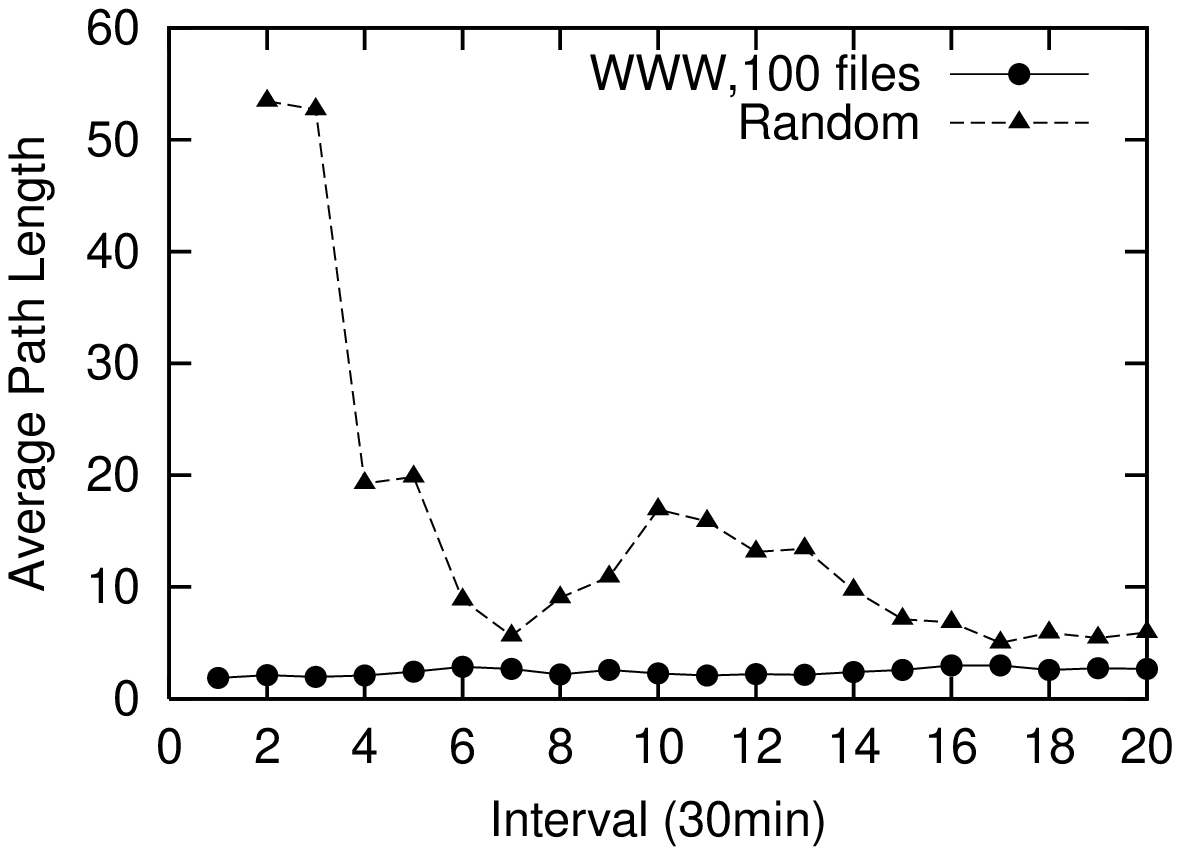}
\caption{Clustering coefficients (left) and average path lengths (right) of
WWW data-sharing graphs and random graphs of same size. Similarity
criterion: 10, respectively 100 shared requests during a half-hour interval.}
\label{fig:web-1800s-sw}
\end{center}
\end{figure}

\begin{figure}[htpb]
\begin{center}
\includegraphics[angle=270,width=1.7in]{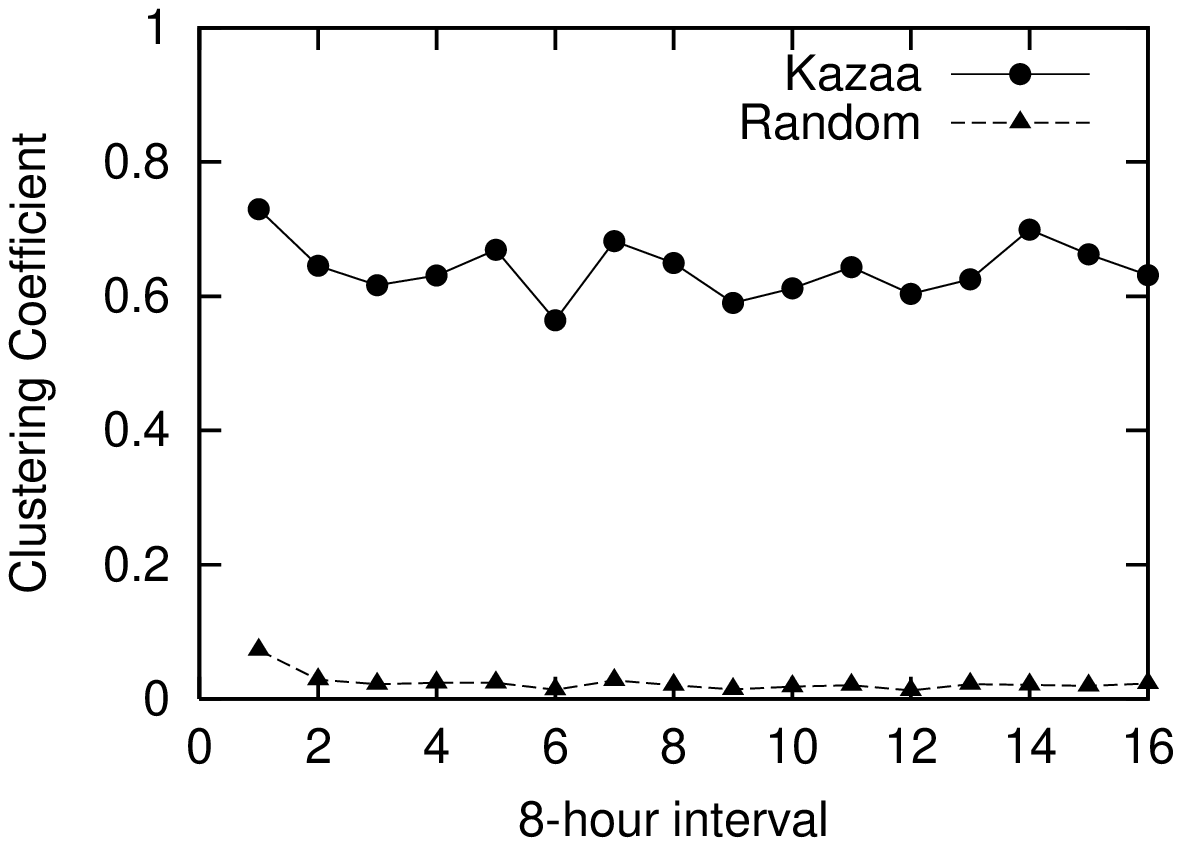}
\includegraphics[angle=270,width=1.7in]{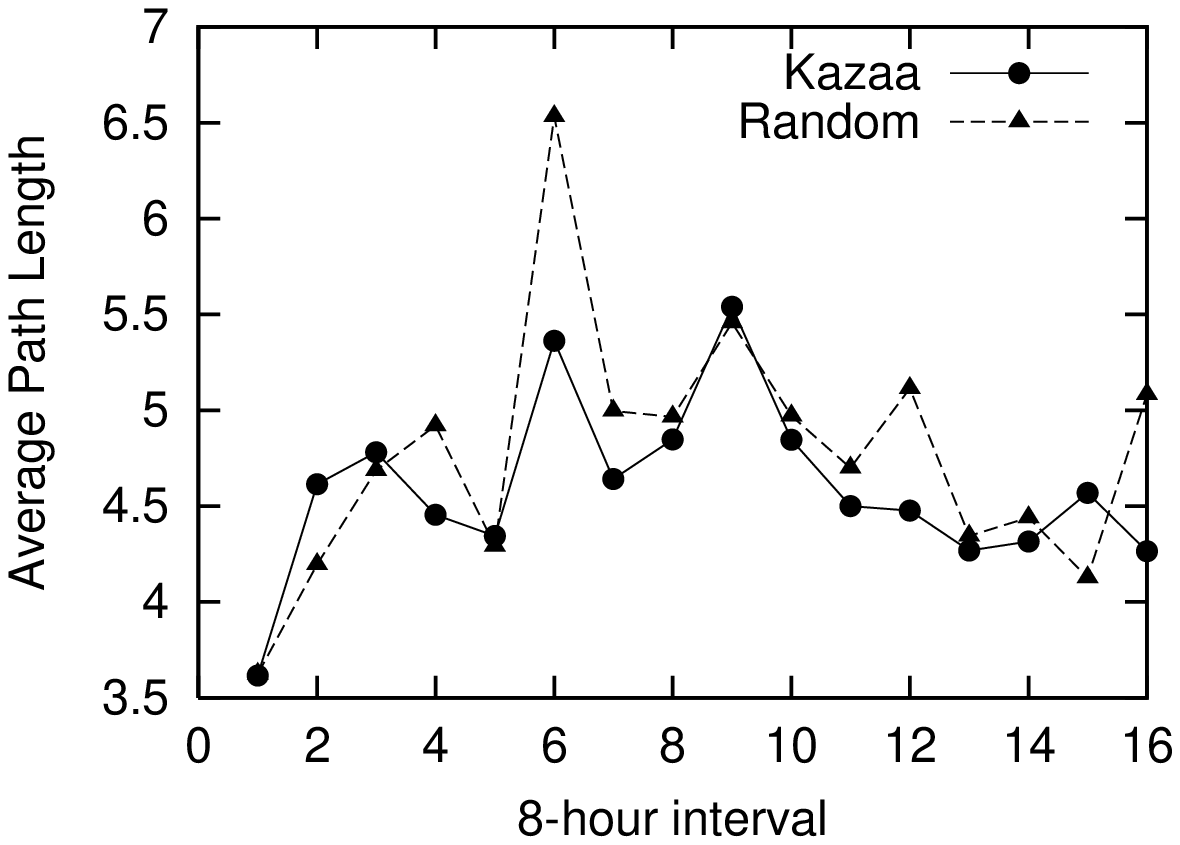}
\caption{Clustering coefficients (left) and average path lengths (right) of
Kazaa data-sharing graphs and random graphs of same size. Similarity
criterion: 2 shared requests during an 8-hour interval.}
\label{fig:kazaa-sw-8h2f}
\end{center}
\end{figure}

\begin{figure}[htbp]
\begin{center}
\includegraphics[width=3.4in]{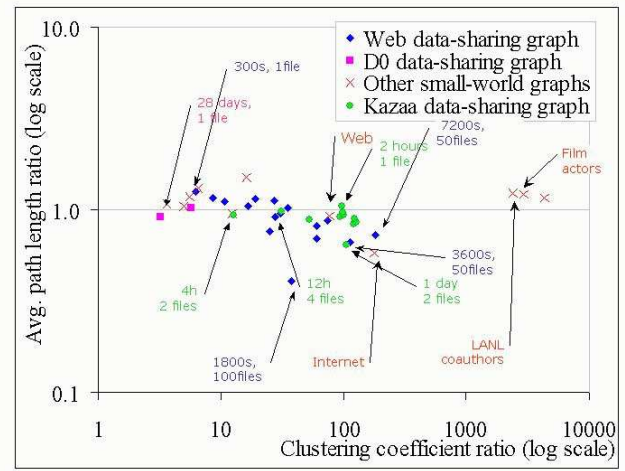}
\caption{Small-world networks: data-sharing graphs and networks
previously documented in the literature as small worlds}
\label{fig:all-sw}
\end{center}
\end{figure}

We discover that data-sharing graphs for the three systems all display 
small-world properties. Figures \ref{fig:d0-cc},
\ref{fig:web-1800s-sw}, and
\ref{fig:kazaa-sw-8h2f} show the small-world patterns---large
clustering coefficient and small average path length---remain constant
over time, for the entire period of our studies. Figure
\ref{fig:all-sw} summarizes the small-world result: it compares some
instances of data-sharing graphs with small-world 
networks already documented in the literature. The axes represent the
ratios of the data-sharing graphs metrics and the same metrics of
random graphs of same size. Notice that most datapoints are
concentrated around $y=1$ (``same average path length'') and $x>10$
(``much larger clustering coefficient'').

We clearly see that data-sharing graphs of various durations and
similarity criteria are small worlds. From the Watts-Strogatz model of
small worlds---as loosely connected collections of highly connected
subgraphs---two significant observations can be drawn. First, well
connected clusters exist; due to the data-sharing graph definition,
these clusters map onto groups of users with shared interests in
files. Second, there is, on average, a small path between any two nodes
in the data-sharing graph: therefore, for example, flooding
with relatively small time-to-live would cover most of the graph.

\section{Human Nature or Zipf's Law?}
\label{sec:HNZL}

We observed small-world patterns in three different file-sharing
communities: a scientific collaboration, the Web, and the Kazaa
peer-to-peer system. Given the variety of our study sample, we could
perhaps generalize this observation to any file-sharing user
community. Thus, we seek to understand what causes these
characteristics in data-sharing graphs and to answer the question:  

\begin{itemize}
\item[\textit{Q3}] \textit{Are the small-world characteristics
consequences of previously 
documented patterns or do they reflect a new observation concerning
user's preferences in data?}  
\end{itemize}


We explore two directions that help us answer the causality
question. In Section \ref{sec:PrefNet} we focus on the definition of
the data-sharing graph and question the large clustering coefficient
as a natural consequence of the graph definition. In Section 
\ref{sec:shuffleDSG} we analyze the influence of well-known patterns
in file access, such as time locality and file popularity
distribution.

\subsection{Affiliation Networks}
\label{sec:PrefNet}

\textit{An affiliation
network} (also called ``a preference network'') is a social network in
which the participants (\textit{actors} in sociology terminology) are
linked by common membership in groups or 
clubs of some kind. Examples 
include scientific collaboration networks (in which actors 
belong to the group of authors of a scientific paper), movie actors
(in which actors belong to the cast of a certain movie), and board
directors (in which actors belong to the same board). 
 
\begin{figure}[htpb]
\begin{center}
\includegraphics[width=1.8in]{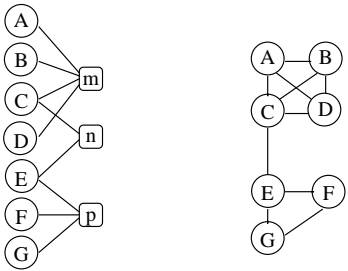}
\caption{A bipartite network (left) and its unipartite projection
(right). Users A-G access files m-p. In the unipartite projection, two 
users are connected if they requested the same file.}
\label{fig:example-bipartite}
\end{center}
\end{figure}

Affiliation networks are therefore bipartite graphs: there are two
types of vertices, for actors and respectively groups, and edges
link nodes of different types only (Figure
\ref{fig:example-bipartite}, left). Affiliation networks are often
represented as unipartite graphs of actors joined by undirected edges
that connect actors in the same group. One observes now that the 
data-sharing graph with one-shared file threshold for the similarity
criterion is such a one-mode projection of a
bipartite affiliation network (Figure \ref{fig:example-bipartite},
right).

These one-mode projections of bipartite graphs have particular
characteristics. Most relevant to this discussion is the clustering
coefficient: inherently, the clustering coefficient is larger in these
graphs than in random graphs of the same size, since the members of a
group will form a complete subgraph in the one-mode
projection. Consequently, our comparison with random graphs, although 
faithful to the Watts-Strogatz definition of small worlds, is
misleading. 

We therefore identified two possible sources of bias in our analysis:
one is the 
implicitly large clustering coefficient of the unimodal affiliation
networks, as just shown. Another is the degree distribution of the
data-sharing graphs which, as in many other real networks, is far from
the Poisson distribution of a random graph (Figures
 \ref{fig:d0-dd}, \ref{fig:web-dd}, and \ref{fig:kazaa-dd}).

Newman et al. \cite{newman02random, newman00-2} propose a model for
random graphs with given degree distributions. These
graphs, therefore, will not be random in the Erd\H{o}s-R\'{e}nyi
sense, but will be random members of a 
class of graphs with a fixed degree distribution. The authors also adapt
their model to affiliation networks and deduce a set of parameters of
their unimodal projection. We use their theoretical model to estimate
the clustering coefficient of unimodal projections of random
affiliation networks of the size and degree distributions as given by 
traces and compare it with the actual values.   

In a bipartite affiliation network, there are two degree distributions:
of actors (to how many groups does an actor belong) and of groups (how many
actors does a group contain). Let us consider a bipartite affiliation
graph of $N$ actors and $M$ groups. Let us name $p_j$ the
probability that an actor is part of exactly $j$ groups and  $q_k$ the probability
that a group consists of exactly $k$ members. In order to easily
compute the average node degree and the clustering coefficient of the
unipartite affiliation network, Newman et al. use three functions $f_0$,
$g_0$, and $G_0$ defined as follows:

\begin{equation}
f_0(x) = \sum_{j=1}^{N}p_jx^j
\end{equation}

\begin{equation}
g_0(x) = \sum_{k=1}^{M}q_kx^k
\end{equation}

\begin{equation}
G_0(x) = f_0(g'_0(x)/g'_0(1))
\end{equation}

The average degree for the actors' one-mode projection of the
affiliation network is: 
\begin{equation}
AvgDegree = G'_0(1)
\end{equation}

And the clustering coefficient is:
\begin{equation}
C = \frac{M}{N}\frac{g'''_0(1)}{G''_0(1)}
\label{eq:CC-bipartite}
\end{equation}

The definition of the clustering coefficient is that of
Eq. \ref{eq:CC2}.  

It is therefore relevant to compare
the clustering coefficient of data-sharing graphs with that given by
Equation \ref{eq:CC-bipartite}. 

\begin{figure}[htpb]
\begin{center}
\includegraphics[angle=270,width=1.7in]{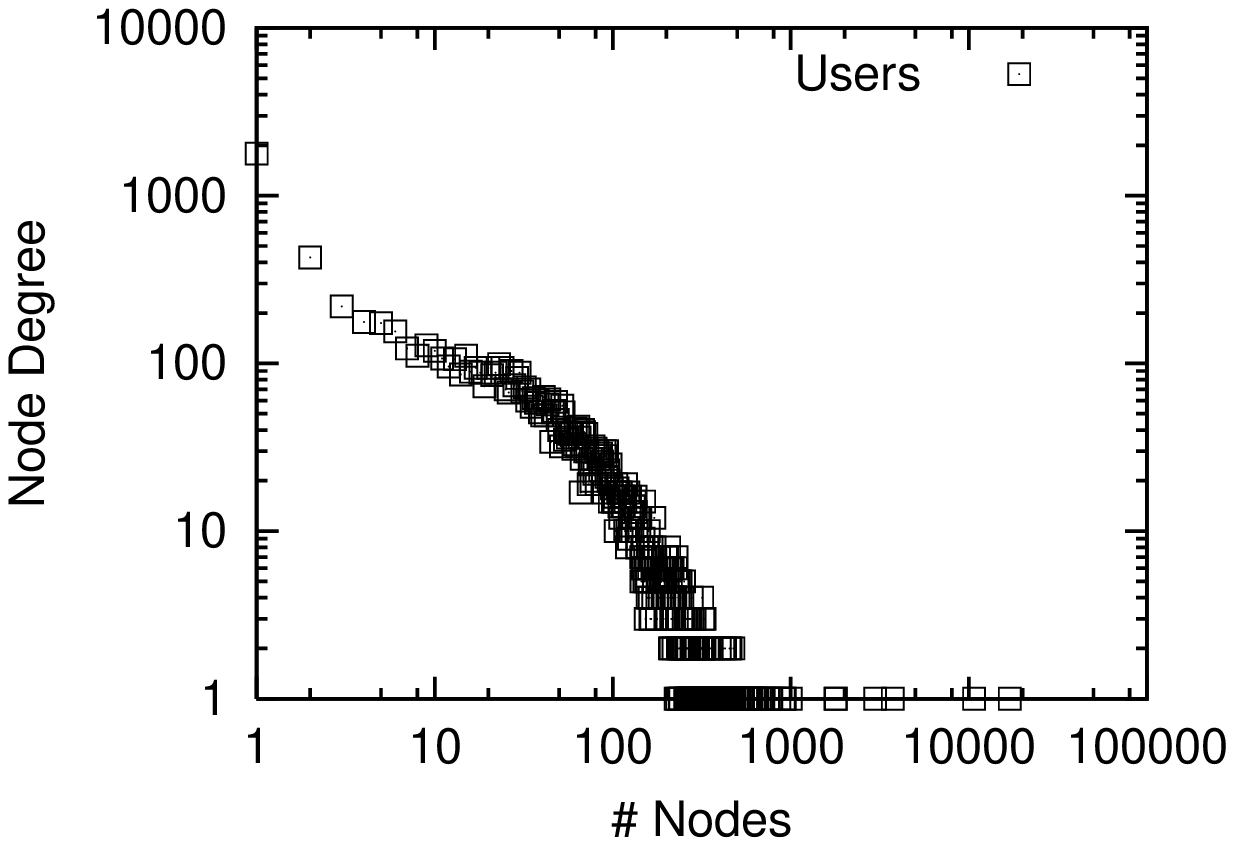}
\includegraphics[angle=270,width=1.7in]{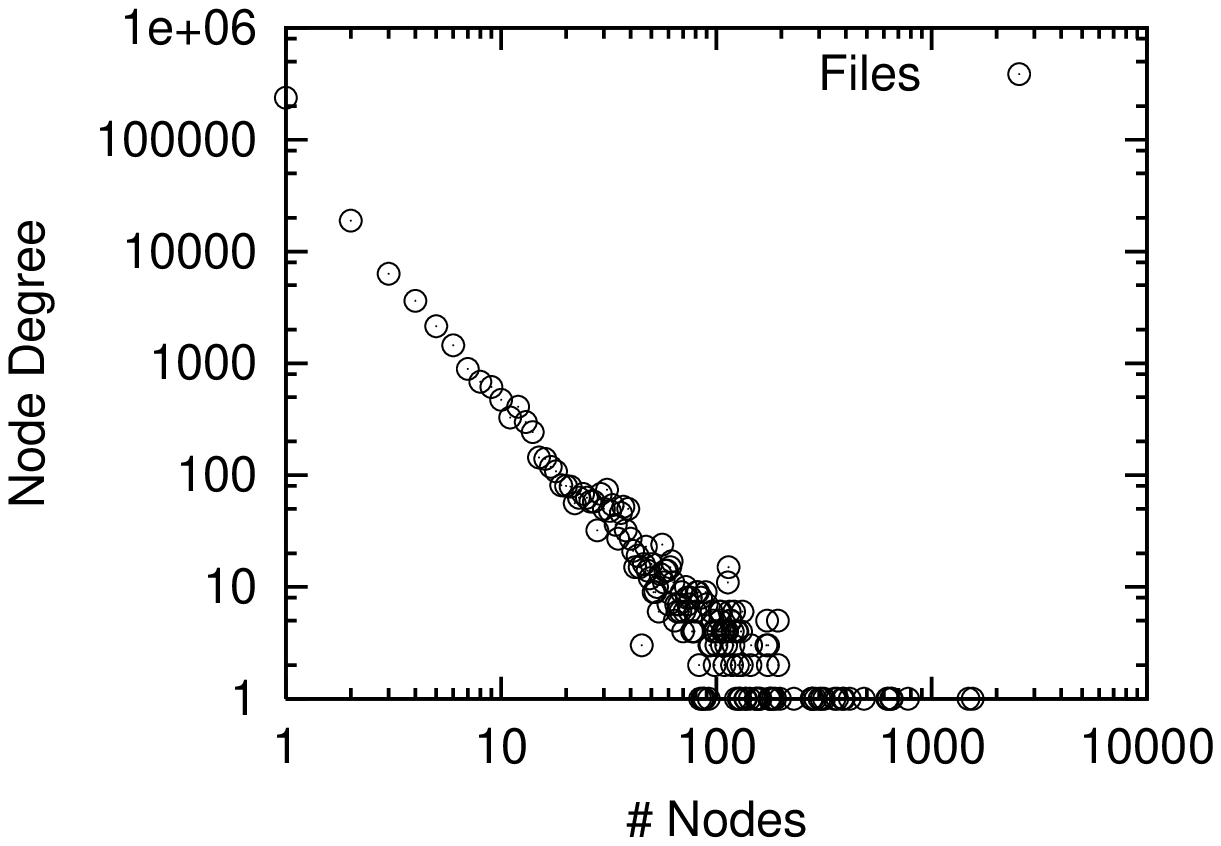}
\caption{Degree distribution of user (left) and file (right) nodes of
a bipartite affiliation network corresponding to a half-hour interval in
the Boeing Web traces.}
\label{fig:web-bipartite-dd}
\end{center}
\end{figure}

Figure \ref{fig:web-bipartite-dd} shows the corresponding values for
the degree distribution $p$ and $q$ (but not normalized: i.e., it
shows the number rather than the percentage of users that requested
exactly $k$ files) in a Web data-sharing graph with a similarity
criterion of one shared request within a half-hour interval. 

Table \ref{table:newmans-model} shows that our intuition was correct:
there is a significant difference 
between the values of measured and modeled parameters. Thus, the large
clustering coefficient is not due to 
the definition of the data-sharing graph as a one-mode projection of
an affiliation network with non-Poisson 
degree distributions. 


\begin{table*}
\begin{center}
\caption{Properties of data-sharing graphs, measured and modeled
as unimodal projection of affiliation networks. Clustering 
coefficient are measured using Eq. \ref{eq:CC2} and modeled using 
Eq. \ref{eq:CC-bipartite}}  

\begin{tabular}{|l|r|r|r|l|r|r|r|}
        \hline
        \hline
&&&& \multicolumn{2}{|c|}{Clustering} &
\multicolumn{2}{|c|}{Average degree}\\
 & Interval & Users  & Files & Theory & Measured & Theory &
Measured\\
        \hline
D0 & 7 days & 74 & 28638 & 0.0006 & 0.65 & 1242.5 & 3.3\\
 & 28 days & 151 & 67742 & 0.0004 & 0.64 & 7589.6 & 6.0\\
        \hline
Web & 2 min & 3385 & 39423 & 0.046 & 0.63 & 50.0 & 22.9\\
 & 30 min & 6757 & 240927 & 0.016 & & 1453.1 & 304.1\\
        \hline
Kazaa & 1 h & 1629 & 3393 & 0.55 & 0.60 & 2.9 & 2.4 \\
 & 8 h & 2497 & 9224 & 0.30 & 0.48 & 9.5 & 8.7\\
        \hline
        \hline
\end{tabular}
\label{table:newmans-model}
\end{center}
\end{table*}


Table \ref{table:newmans-model} leads to two observations. First,
the actual clustering coefficient in the data-sharing graphs is always
larger than predicted and the average degree is always smaller than
predicted. An interesting new question emerges: what is the
explanation for these (sometimes significant) differences? One
possible explanation is that user requests for files are not random:
their preferences are limited to a set of files, which explains the
actual average degree being smaller than predicted. A rigorous
understanding of this problem is left for future work.

A second observation is that we can perhaps compare the file sharing
in the three communities by comparing their distance from the
theoretical model. We see that the Kazaa data-sharing graphs are
the closest to the theoretical model and the D0 graphs are very
different from their corresponding model. This is different from the
comparison with the  Erd\H{o}s-R\'{e}nyi random
graphs (Table \ref{table:all}). The cause of this difference
and the significance of this observation remain to be studied
in the future.

\subsection{Influences of Zipf's Law and Time and Space Locality}
\label{sec:shuffleDSG}

Event frequency has been shown to follow a Zipf distribution in many
systems, from word occurrences in English  
and in monkey-typing texts to city population. It is also present in two
of the three cases we  analyze: the Web and Kazaa. Other 
patterns characteristic to data access systems include time locality, 
in which an item is more popular (and possibly requested by multiple
users) during a limited interval and temporal user activity, meaning
that users are 
not uniformly active during a period, but follow some patterns (for
example, downloading more music files during weekends or holidays
\cite{ripeanu02mapping}). Thus, we ask:

\begin{itemize}
\item [\textit{Q4}] \textit{Are the patterns we identified in the
data-sharing graph, especially 
the large clustering coefficient, an inherent consequence of these
well-known behaviors?}  
\end{itemize}

To answer this question, we generate random traces that preserve the
documented characteristics but break the user-request
association. From these synthetic traces, we build the resulting 
data-sharing graphs, and analyze and compare their properties with
those resulting from the real traces.

\subsubsection{Synthetic Traces}
\label{sec:synthetic-traces}

The core of our traces is a triplet of user ID, item requested
and request time. Figure \ref{fig:urt-relationship} identifies the
following correlations in traces, some of which we 
want to preserve in the synthetic traces:

\begin{figure}[htpb]
\begin{center}
\includegraphics[width=2.3in]{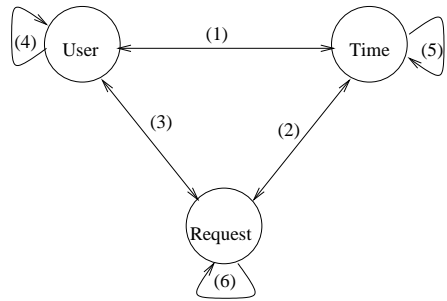}
\caption{The relations between users, their requests, and their
request times determine observed patterns like Zipf frequency of
requests or time locality.}
\label{fig:urt-relationship}
\end{center}
\end{figure}

\begin{enumerate}
\item[(1)] User--Time: User's activity varies over time: for example,
in the D0 traces, some users accessed data only in May. 

\item[(2)] Request--Time: Items may be more popular during some
intervals: for example, news sites are more popular in the
morning. 

\item[(3)] User--Request: This is the key to user's preferences. By
breaking this relationship and randomly recreating it, we can analyze
the effect of user preferences on the properties of the data-sharing
graph. 

\item[(4)] User: The number of items requested per user over the
entire interval studied may be relevant, as some users are more active
than others (see Figures \ref{fig:web-reqs-per-user} left for the Web
traces).  

\item[(5)] Time: The time of the day (or in our case, of the periods
studied) is relevant, as the Web traces show (the peak in Figure
\ref{fig:web-reqs-per-user} right).

\item[(6)] Request: This is item popularity: number of requests for
the same item. 
\end{enumerate}

Our aim is to break the relationship (3), which implicitly requires
the break of (1), (2), or both. We also want to preserve relationships
(4), (5), and (6). 

One can picture the traces as a $R\times3$ matrix, in which $R$ is the
number of requests in that trace and the three columns correspond to users,
files requested, and request times, respectively. Now imagine the
we shuffle the users column while the other two are kept unchanged: this
breaks relations (3) and (1). If the requests column is shuffled,
relations (3) and (2) are broken. If both user and request columns
are shuffled, then relations (1), (2), and (3) are broken. In all
cases, (4), (5), and (6) are maintained faithful to the real behavior:
that is, users ask the same number of requests (4); the times
when requests are sent are the same (5); and the same requests
are asked and repeated the same number of times (6). 

We generated synthetic traces in three ways, as presented above:
\begin{enumerate}
\item[ST1:] No correlation related to time is maintained: break
relations (1), (2), and (3).
\item[ST2:] Maintain the request times as in the real traces: break
relations (1) and (3).
\item[ST3:] Maintain the user's activity over time as in the real
traces: break (2) and (3).  
\end{enumerate}

\subsubsection{Properties of Synthetic Data-Sharing Graphs}
\label{sec:synthetic-dsg}

\begin{figure}[htbp]
\begin{center}
\includegraphics[angle=270,width=1.7in]{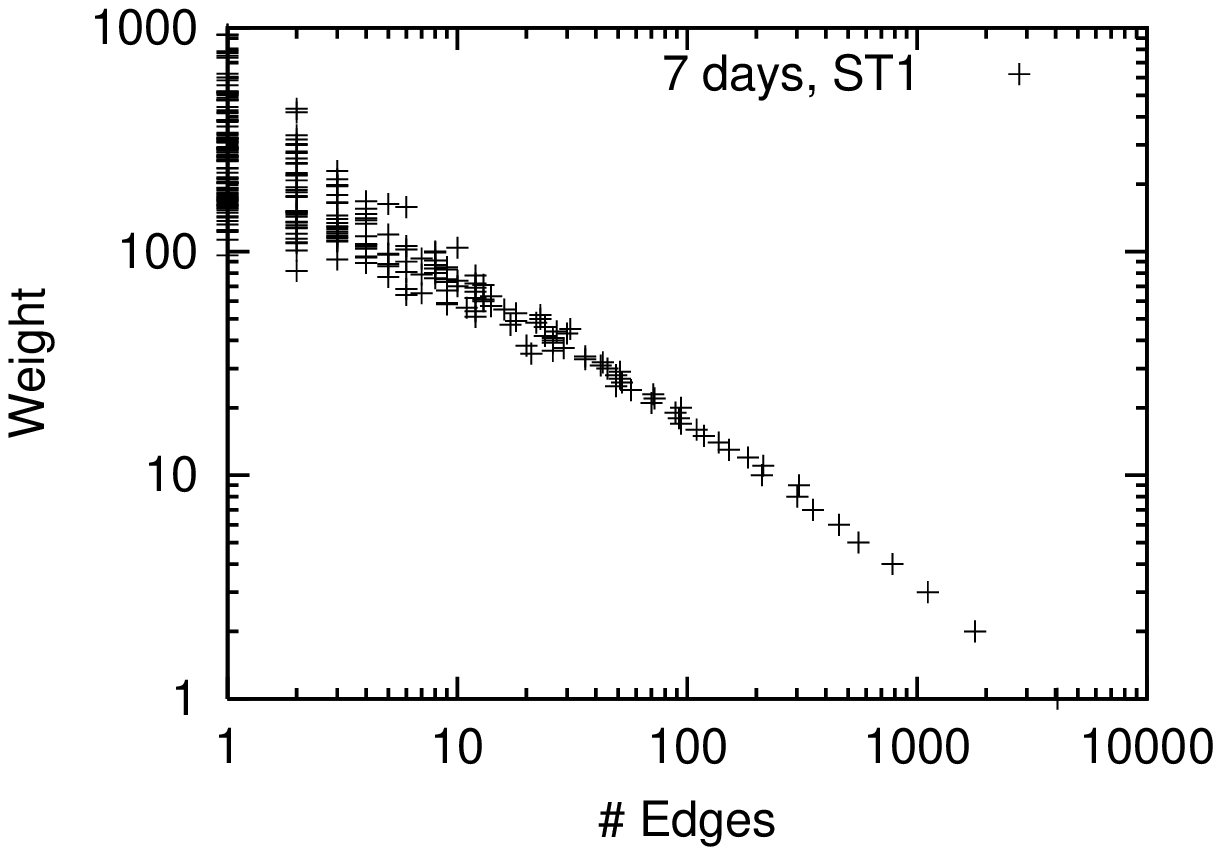}
\includegraphics[angle=270,width=1.7in]{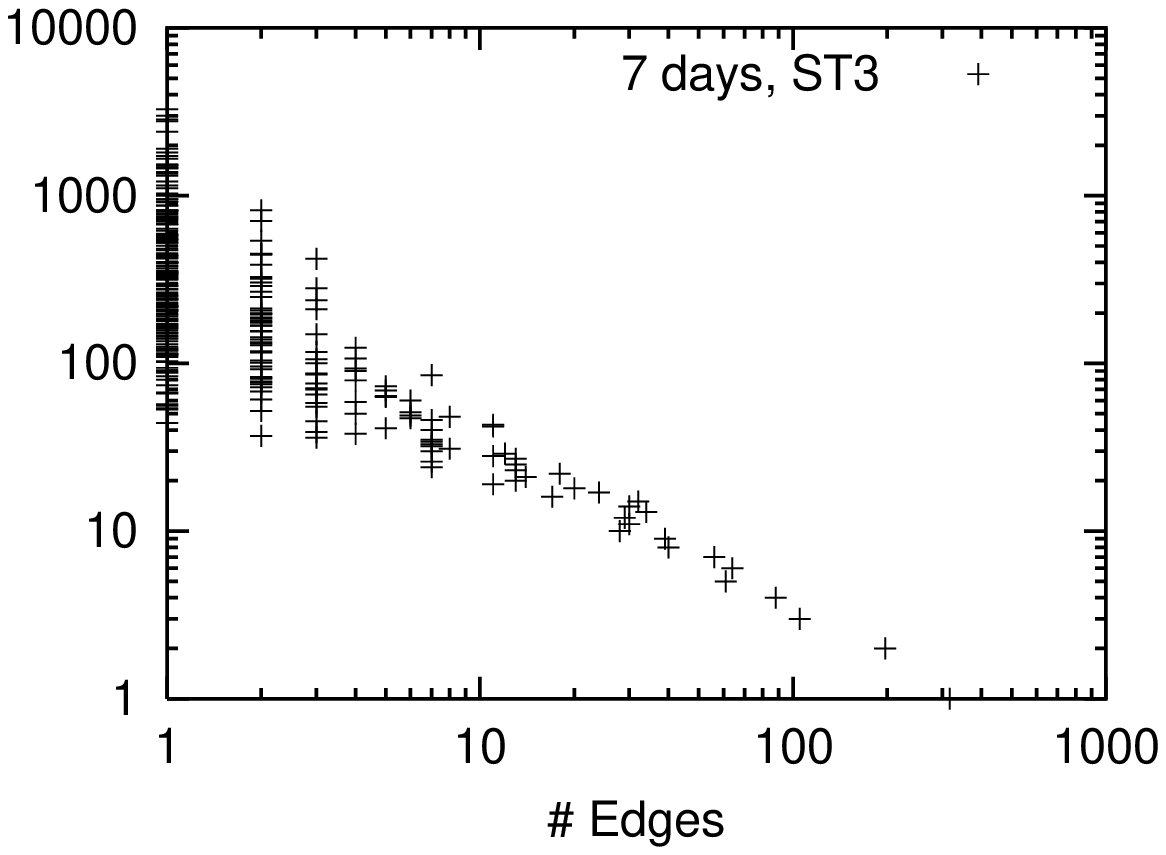}
\caption{Distribution of weights in the synthetic data-sharing graphs
built from shuffling the D0 traces.}
\label{fig:synth-d0-wd}
\end{center}
\end{figure}

\begin{figure}[htbp]
\begin{center}
\includegraphics[angle=270,width=1.7in]{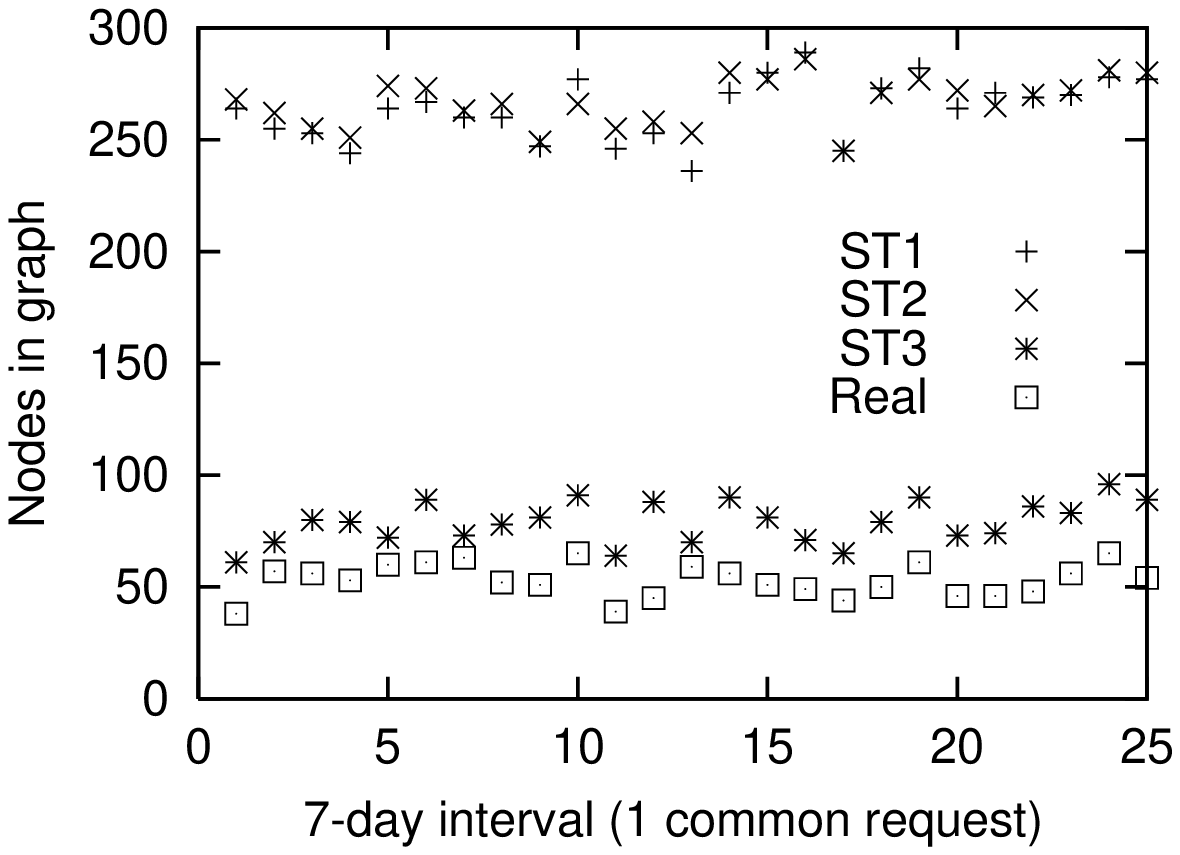}  
\includegraphics[angle=270,width=1.7in]{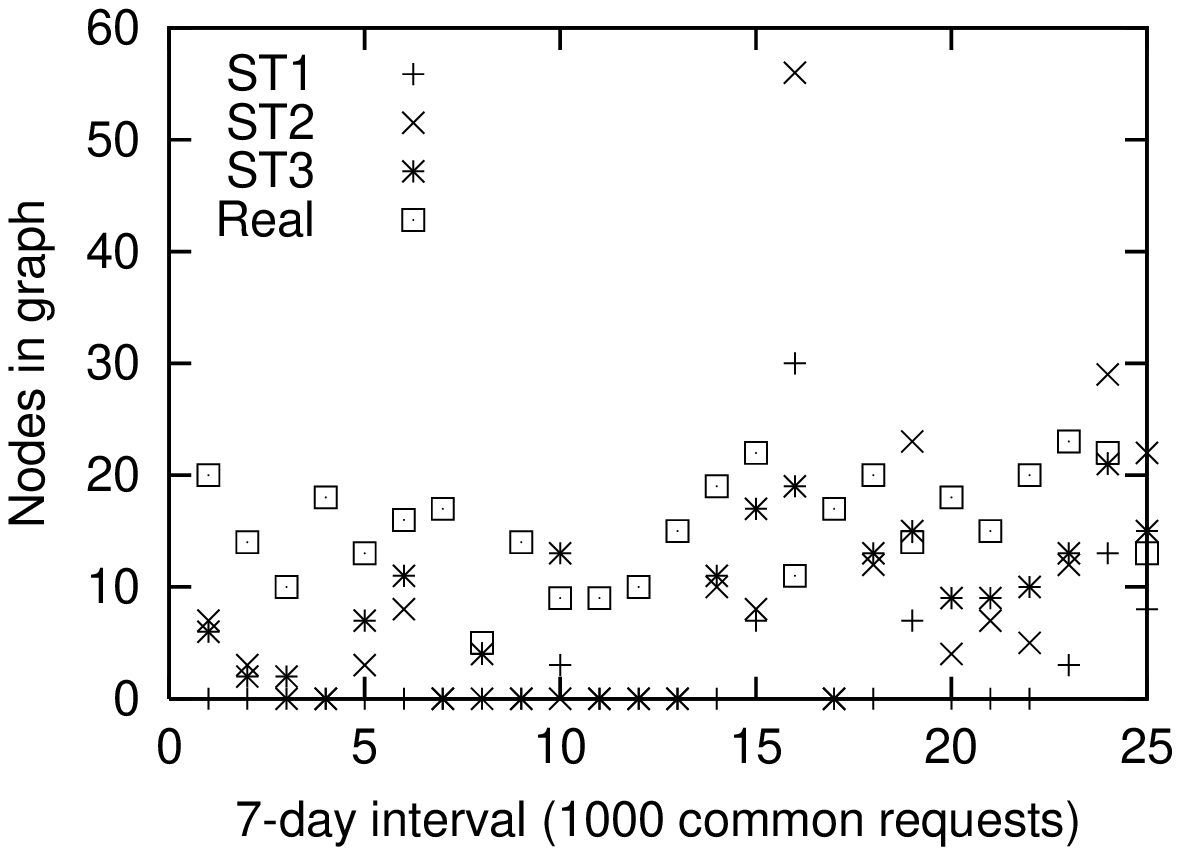}
\caption{Number of nodes in data-sharing graphs in real and synthetic
D0 traces}
\label{fig:d0-nnodes}
\end{center}
\end{figure}

Three characteristics of the synthetic data-sharing graphs are
relevant to our study. First, the number of nodes in synthetic graphs
is significantly different than in their corresponding real
graphs (``corresponding'' in terms of similarity criterion and
time). On the one hand, the synthetic data-sharing graphs for
which user activity in time (relation (1)) is not preserved have
a significantly larger number of nodes. Even when the user activity in
time is preserved (as in the ST3 case), the number of nodes is
larger: this is because in the real data-sharing graphs, we ignored the
isolated nodes and in the synthetic graphs there are no isolated nodes.   
On the other hand, when the similarity criterion varies to a large
number of common requests (say, 100 in the D0 case, Figure
\ref{fig:d0-nnodes}), the synthetic graphs are much smaller or even
disappear. This behavior is explained 
by the distribution of weights in the synthetic graphs (Figure 
\ref{fig:synth-d0-wd}): compared to the real graphs (Figure
\ref{fig:d0-weightdistrib}), there are many more edges with small
weights. The median weight in the real D0 data-sharing graphs is 356
and the average is 657.9, while for synthetic graphs the median is 137
(185 for ST3) and the average is 13.8 (75.6 for ST3).

Second, the synthetic data-sharing graphs are always connected
(unlike real graphs, that always have multiple connected components,
as shown in Table \ref{table:all}). Even for similarity criteria with
large number of common requests the synthetic graphs remain connected. This
behavior is due to the uniform distribution of requests per user in
the case of synthetic traces, which is obviously not true in the real
case. 

\begin{figure}[htbp]
\begin{center}
\includegraphics[angle=270,width=2.4in]{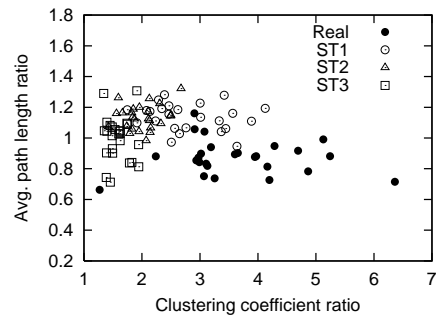}
\caption{Comparison of the small-world data-sharing graphs as resulted
from the real and synthetic D0 traces. }
\label{fig:synth-sw-d0}
\end{center}
\end{figure}

Third, the synthetic data-sharing graphs are ``less'' small worlds
than their corresponding real graphs: the ratio between the clustering
coefficients is smaller and the ratio between average path lengths is
larger than in real data-sharing graph (Figure
\ref{fig:synth-sw-d0}). However, these 
differences are not major: the synthetic data-sharing graphs would
perhaps pass as small worlds.

These results show that user preferences for files have
significant influence on the data-sharing graphs: their properties are
not induced (solely) by user-independent trace characteristics, but human
nature has some impact. So perhaps the answer to this section title
(``Human nature or Zipf's law?'') is ``Both''. However, it seems that
identifying small-world 
properties is not a sufficient metric to characterize the natural
interest-based clustering of users: we might need a metric of how
small world a small-world data-sharing graph is. This problem remains to be
studied further in the future.  

\section{Small-World Data-Sharing Graph: Significance for
Mechanism Design}
\label{sec:DSG-uses}

It is interesting to notice that the structure we call the
data-sharing graph can be applied at various levels and granularities
in a computing system. We looked at relationships that form at the
file access level, but intuitively similar patterns could be found at
finer granularity, such as access to same memory locations or access to same 
items in a database. For example, a recent article
\cite{sendag03address} investigates  the correlation of program
addresses that reference the same data and shows that these
correlations can be used to eliminate load misses and partial hits. 

At a higher level, the data-sharing graph can identify the
structure of an organization---based on the applications its members
use, for example---by identifying interest-based clusters of users and
then use this information to optimize an organization's
infrastructure, such as servers or network topology.

In this section we focus on implications for mechanism design of the
data-sharing graph from two perspective: its structure (definition)
and its small-world properties. We stress that these are
untested but promising ideas for future work.


\subsection{Relevance of the Data-Sharing Graph Structure}

Some recommender systems have a similar flavor to the data-sharing
graph. ReferralWeb \cite{kautz97referralweb} 
attempts to uncover existing social networks to create a referral chain
of named individuals. It does this by inferring social relationships
from web pages, such as co-authorship, research groups and interests,
co-participation in discussion panels, etc. This social network is
then used to identify experts and to guide searches around them. 

Sripanidkulchai et. al came close to the intuition of the
data-sharing graph in their Infocom 2003 article
\cite{sripanidkulchai03efficient}: they improve 
Gnutella's flooding-based mechanism by inserting and exploiting
interest-based shortcuts between peers. Interest-based shortcuts
connect a peer to peers who provided data in the past. This
is slightly different from our case, where an edge in the data-sharing
graph connects peers that requested the same data. However, the two graphs
are likely to overlap significantly if peers store data of their own
interest. Our study distinguishes by its independence from any
underlying infrastructure (in this case, the distribution of data on
peers and the location mechanism) and gives a theoretical explanation
of the performance improvements in \cite{sripanidkulchai03efficient}.

The data-sharing graph can be exploited for a variety of decentralized
file management mechanisms in resource-sharing systems (such as
peer-to-peer or Grids). 
\begin{itemize}
\item In a writable file-sharing system,
keeping track of which peers recently requested a file facilitates the
efficient propagation of updates in a fully decentralized, 
self-organizing fashion (a similar idea is explored in
\cite{saito02taming}).  

\item In large-scale, unreliable, dynamic peer-to-peer systems file
replication may be used to insure data availability
\cite{ranganathan02improving} and transfer performance. The
data-sharing graph may suggest where to place replicas 
closer to the nodes that access them. Similarly, it may be useful for
dynamic distributed storage: if files cannot be stored entirely
on a node, then they can be partitioned among the nodes
that are interested in that file.  

\item In a peer-to-peer computing scenario, the relationships between
users who requested the same files can be exploited for job
management. If nodes store and share recently downloaded files,
they become good candidates for running jobs that take those files as
input. This can be used for scheduling, migrating or replicating
data-intensive jobs.
\end{itemize}

\subsection{Relevance of Small-World Characteristics}

The idea underlying the data-sharing graph was first presented in
\cite{iamnitchi02locating} as a challenge to design a file-location
mechanism that exploits the small-world characteristics of a
file-sharing community. Meanwhile we completed the design and
evaluation of a mechanism that dynamically identifies 
interest-based clusters, disseminates location information in groups
of interested users, and propagates requests among clusters
\cite{iamnitchi-unpublished}. Its 
strengths come from mirroring and adapting to changes in user's
behavior. File insertion and deletion are low cost, which makes it a
good candidate for scientific collaborations, where use of files leads
to creation of new files. 

\section{Summary}
\label{sec:summary}

This article reveals a predominant pattern in diverse file-sharing
communities, from scientific communities to the Web and file-swapping
peer-to-peer systems. This pattern is brought to light by a structure
we propose and that we call ``data-sharing graph''. This structure
captures the relationships that form between users who are interested in
the same files. We present properties of data-sharing graphs from
three communities. These properties are relevant to and might inspire the
design of a new style of mechanisms in peer-to-peer systems,
mechanisms that take into account, adapt to, and exploit user's
behavior. We also sketch some mechanisms that could benefit from the
data-sharing graph and its small-world properties.


\end{document}